\documentclass[11pt,a4paper]{article}
\pdfoutput=1
\usepackage{jheppub}
\usepackage{amsfonts}
\usepackage{amssymb}
\usepackage{bbm}
\usepackage{graphicx}
\usepackage{caption}
\usepackage{subcaption}

\usepackage[normalem]{ulem}

\def\be{\begin{equation}}
\def\ee{\end{equation}}
\def\bea{\begin{eqnarray}}
\def\eea{\end{eqnarray}}

\newcommand{\nn}{\nonumber}

\newcommand\diff{\mathrm{d}}

\numberwithin{equation}{section}       % equation numbers in each section

\def\rme{{\rm e}}
\def\elpot{\varphi}

\newcommand{\ii}{\mathrm{i}}

\renewcommand{\=}{\,= \,}

\renewcommand{\d}{\delta}

\newcommand{\s}{\tau}
\renewcommand{\t}{\sigma}

\newcommand\om{\omega}
\renewcommand{\nn}{\nonumber}

\renewcommand{\Re}{\text{Re}}
\renewcommand{\Im}{\text{Im}}

\newcommand{\ve}{\varepsilon}
\newcommand{\CN}{\mathcal{N}}

\newcommand{\CH}{\mathcal{H}}

\newcommand{\p}{\partial}

\newcommand{\CI}{\mathcal I}
\newcommand{\CF}{\mathcal F}

\newcommand{\IZ}{\mathbb Z}
\newcommand{\IR}{\mathbb R}

\newcommand{\ndt}{\noindent}

\renewcommand{\i}{{\rm i}}

\newcommand{\Ge}{\Gamma_\text{e}}

\usepackage{array}

\title{The asymptotic growth of states of the 4d $\CN=1$ superconformal index}

\author[a]{Alejandro Cabo-Bizet,}
\emailAdd{alejandro.cabo\_bizet@kcl.ac.uk}
\author[b]{Davide Cassani,}
\emailAdd{davide.cassani@pd.infn.it}
\author[a,\dagger]{Dario Martelli,\footnotetext[2]{On leave at the Galileo Galilei Institute, Largo Enrico Fermi, 2, 50125 Firenze, Italy.}}
\emailAdd{dario.martelli@kcl.ac.uk}
\author[a]{Sameer Murthy}
\emailAdd{sameer.murthy@kcl.ac.uk}
\affiliation[a]{Department of Mathematics, King's College London,\\
The Strand, London WC2R 2LS, U.K.}
\affiliation[b]{\it INFN, Sezione di Padova, \\
Via Marzolo 8, 35131 Padova, Italy}

\abstract{We show that the superconformal index of~$\CN=1$ superconformal field theories in four dimensions
has an asymptotic growth of states which is exponential in the charges. 
Our analysis holds in a Cardy-like limit of large charges, for which the index is dominated by small values of chemical potentials. 
In this limit we find the saddle points of the integral that defines the superconformal index using two different methods. 
One method, valid for finite~$N$,  is to first take the Cardy-like limit and then find the saddle points. 
The other method is to analyze the saddle points at large~$N$ and then take the Cardy-like limit. 
The result of both analyses is that the asymptotic growth of states of the superconformal index exactly agrees with the 
Bekenstein-Hawking entropy of supersymmetric black holes in the dual~AdS$_5$ theory.
}

\usepackage{amsmath}

\begin{document}
 
\maketitle

%%%%%%%%%%%%%%%%%%%%%%%%%%%%%%%%%%%%%%%%%%%%%%%%%%%%%%%%%%%%%%%%%%%%%%%%%%%%%%%%%%%%%%

\section{Introduction}\label{sec:Intro}

Recently there has been renewed interest in the subject of the entropy of supersymmetric black holes 
in~AdS$_5$~\cite{Cabo-Bizet:2018ehj,Choi:2018hmj, Choi:2018vbz, Benini:2018ywd,Honda:2019cio, ArabiArdehali:2019tdm},
inspired by~\cite{Hosseini:2017mds,Benini:2015noa,Benini:2015eyy}. 
Such black holes were discovered as solutions to minimal gauged supergravity in five dimensions~\cite{Gutowski:2004ez} 
and generalized in~\cite{Gutowski:2004yv,Chong:2005hr,Chong:2005da,Kunduri:2006ek}.
They have a Bekenstein-Hawking entropy~$S_\text{BH} = \frac{A}{4G_5}$ where~$A$ is the horizon area and~$G_5$ is Newton's 
constant in five dimensions. 
The holographical dual boundary theories are four-dimensional~$\CN=1$ superconformal field theories (SCFTs), the most studied case
being~$\CN=4$ SYM theory with~$SU(N)$ gauge group. The existence of the above-mentioned black holes with large entropy 
implies that the corresponding supersymmetric ensemble of states in the boundary theory should have an exponentially large 
number of states~$\sim\mathcal{O}(\rme^{N^2})$ 
Trying to identify these states, however, led to interesting puzzles~\cite{Kinney:2005ej}. 

The renewal of interest in this subject was sparked in large measure by the insightful observation of~\cite{Hosseini:2017mds} 
which recast the entropy of the black holes as an extremization problem as follows. Consider a supersymmetric black 
hole in AdS$_5$ carrying angular momenta~$J_{1,2}$ and~$R$-charge~$Q$. The Bekenstein-Hawking entropy of 
this black hole is then the Legendre transform 
\be \label{Legendretransf}
S(J_1,J_2,Q) \= -I(\omega_1,\omega_2,\elpot) - \omega_1  J_1 - \omega_2  J_2 - \elpot \, Q \ ,
\ee
of the function (here~$g$ is the inverse AdS radius), 
\be  \label{Iresult}
I = \frac{2\pi}{27 G_5 g^3}\frac{\elpot^3}{\omega_1\omega_2}\ ,
\ee
under the constraint\footnote{The black hole entropy is also reproduced by~$n_0=+1$, and there is a symmetry between
these two choices in that all the equations below can be modified appropriately in order to hold for this choice.} 
\be \label{Constraint}
 \om_1+ \om_2 - 2 \elpot  \= 2 \pi \i \, n_0 \,, \quad n_0 \= - 1\,.
\ee
This observation suggested the existence of a thermodynamic principle which leads to the above extremization formula. 
Finding such a principle for supersymmetric black holes, however, is a subtle issue. One reason is that 
the chemical potentials\footnote{The chemical potentials $\omega_1,\omega_2,\varphi$ are obtained from a limit of the chemical potentials
  $\Omega_1,\Omega_2,\Phi$. We refer to \cite{Cabo-Bizet:2018ehj} for details.}  $\Omega_1,\Omega_2,\Phi$ conjugated to $J_1,J_2,Q$, get frozen to specific 
values\footnote{Specifically, the chemical potentials dual to the angular momenta takes the value~$\Omega^*_{1,2}=1$ and the potential  dual to the R-charge takes the value~$\Phi^*=\frac32$ as~$\beta \to \infty$~\cite{Cabo-Bizet:2018ehj}.}---independent 
of the charges---in the limit of zero temperature and at a naive level there is nothing left to vary
or extremize. Another puzzling feature of this formula  
is the nature of the constraint~\eqref{Constraint}, as the supersymmetric values 
of the chemical potentials obey the real constraint $\Omega^*_{1}+\Omega^*_{2}-2\Phi^*=-1$.

These puzzles were solved in~\cite{Cabo-Bizet:2018ehj}.
The idea is to deform the supersymmetric black hole solution to a family of non-extremal 
solutions that preserve manifest supersymmetry at the cost of allowing for complex field configurations in the Euclidean theory.
The chemical potentials along this family vary around the frozen on-shell values, and we can consider the thermodynamics 
of the fluctuations. 
The family of solutions have a Euclidean cigar-like geometry 
in which the horizon caps off at a finite distance in the interior. In order to preserve
smoothness on the cigar as well as supersymmetry one has to turn on a background value of the  
R-symmetry gauge field such that the Wilson loop around the circle at infinity is non-zero and the Killing spinor is anti-periodic. This background value of the 
gauge field leads precisely to the above constraint~\eqref{Constraint} and to the free energy~\eqref{Iresult}, that we then extremize.
The black hole entropy is recovered by the extremum value in the limit that we reach the original black hole solution.

In this paper we turn to the dual boundary problem in detail. 
We shall try to point out relations with recent papers dealing with this problem as we go along. 
The holographic dual boundary observable is a partition function of the~$\CN=1$ SCFT on~$S^3 \times S^1$
twisted by chemical potentials~$\omega_1,\omega_2$ for the two angular momenta on~$S^3$, and the 
chemical potential~$\elpot$ for the R-charge of the $\CN=1$ SCFT.
It is convenient to present the field theory discussion in the variables~$\omega_1=2 \pi \i \t$,  $\omega_2=2 \pi \i \s$.
The values of~$\Re(\om_{1,2})<0$ for the black hole imply that~$\t$, $\s$ live in the complex upper half-plane.
Supersymmetry of the background implies that the chemical potentials obey the constraint 
\be \label{ConstraintFT}
\t+\, \s - \frac{\elpot}{\pi i }  \= n_0 \,, \quad n_0 \in \mathbb{Z} \,,
\ee
which mirrors the constraint~\eqref{Constraint}.
In~\cite{Cabo-Bizet:2018ehj}, we obtained the following formula for this partition function of an arbitrary $\CN=1$ SCFT:
\be \label{ZFIsplit}
Z (\t,\s,{\elpot})\= \rme^{-\mathcal{F}(\t,\s,{\elpot})} \,
\mathcal{I}(\t,\s,{\elpot}) \,,
\ee
where the prefactor~$\CF$ is related to the supersymmetric Casimir energy~\cite{Assel:2014paa,Assel:2015nca}, 
and~$\CI$ is essentially the Hamiltonian index.
More precisely, when the R-symmetry Wilson line has 
background value~$n_0$ the functional integral translates to the following trace \cite{Cabo-Bizet:2018ehj}, 
\be
\label{IIndef}
\CI (\t,\s;n_0) 
\=  \, {\rm Tr}_{\CH_\text{phys}}\,  \rme^{\pi i (n_0+1) F} \, \rme^{-\beta \{\mathcal{Q},\overline{\mathcal{Q}}\}+
2 \pi \i \t  J_1 + 2 \pi \i \s  J_2 + \elpot \, Q } \,,
\ee
where the three potentials are constrained by~\eqref{ConstraintFT}.
Upon solving the constraint for~$\elpot$ 
in terms of~$\t, \s$, one can write the partition function and the index as functions of~$\t, \s$ and the 
integer~$n_0$.
In particular, for $n_0=\pm1$ the explicit dependence from the fermion number $F$ drops out and 
this takes the form
\be
\CI(\t, \s;\pm 1) \; = \;  
{\rm Tr}_{\CH_\text{phys}}\,  
\rme^{\pi \ii Q}\, \rme^{-\beta \{\mathcal{Q},\bar{\mathcal{Q}}\}  
+2 \pi \i \t (J_1+\frac{1}{2}Q)+2 \pi \i \s (J_2+\frac{1}{2}Q)} \,.
\ee
As discussed in~\cite{Cabo-Bizet:2018ehj}, 
the~$n_0$-dependence in~\eqref{IIndef} can be completely absorbed in a shift of one
of the chemical potentials, say~$\t$, using the spin-statistics theorem, so that
\be\label{RelInd0}
\CI(\t, \s;n_0) \= \CI(\t- n_0,\s;0) \,.
\ee
Note that~$\CI(\t, \s;0)$ is \emph{not} invariant under $\sigma\to\sigma-n_0$, $n_0 \in \IZ$,  
since the R-charges are not necessarily integers for generic~${\cal N}=1$ theories.
The right-hand side of~\eqref{RelInd0} is the familiar  Hamiltonian definition of the superconformal index~\cite{Romelsberger:2005eg,Kinney:2005ej}
\be \label{InI0rel}
\CI(\t, \s;0) 
\=  \, {\rm Tr}_{\CH_\text{phys}}\,  (-1)^F \rme^{-\beta \{\mathcal{Q},\bar{\mathcal{Q}}\}  
+2 \pi \i \t (J_1+\frac{1}{2}Q)+2 \pi \i \s (J_2+\frac{1}{2}Q)} \,.
\ee
By the usual argument that bosonic and 
fermionic states appear in pairs for a given non-zero value of~$\{\mathcal{Q},\bar{\mathcal{Q}} \, \}$,
it is clear that the index~\eqref{InI0rel}, and therefore~\eqref{IIndef}, is protected, and in particular independent of~$\beta$.

Denoting the values of the (generically non-integer) charges~$J_1+\frac12 Q$, $J_2+\frac12 Q$ by~$n_1$, $n_2$, 
respectively, we can expand the index as
\be \label{inddeg}
\CI(\t, \s;n_0) \= \sum_{n_1,n_2} d(n_1,n_2; n_0) \, \rme^{2 \pi \i (\t n_1 + \s n_2)} \,.
\ee 
We shall refer to~$d(n_1,n_2)$ as the indexed degeneracy, or sometimes simply degeneracy, 
of states for a given set of charges labelled by~$(n_1,n_2)$.
It is clear from the trace formula~\eqref{InI0rel} that a constant shift of the chemical potentials only 
changes the phase of the indexed degeneracies and, in particular, 
\be
|d(n_1,n_2; n_0)| \= |d(n_1,n_2; 0)| \,.
\ee
We are interested in calculating the growth of these degeneracies~$d$ as a function of the charges and, relatedly,
the behavior of the index~$\CI$ as a function of its arguments. The relation between these two is that of 
a change of statistical ensemble. 
The holographic dual to the full AdS$_5$ black hole geometry is the canonical ensemble in which the chemical 
potentials are fixed. In this case the input from the supergravity analysis of~\cite{Cabo-Bizet:2018ehj} 
would fix $n_0=-1$ as discussed above. 
On the other hand, one may be interested in the microcanonical ensemble in which all the charges are held fixed. 
In the bulk this means zooming in to the near-horizon of the black hole and studying the corresponding AdS$_2$
theory, as has been emphasized in~\cite{Sen:2008yk}.\footnote{One could associate the words ``flowing from a 4d 
UV theory to a 1d IR theory" to this change of ensemble, as the boundary behavior in~AdS$_2$ is different from 
higher-dimensional~AdS, but making this precise even in simple 2d examples is subtle~\cite{Murthy:2009dq}.} 
In the dual SCFT, the problem is to study the degeneracy of states for fixed values of charges in the field theory, 
which is given by the inverse Laplace transform of the index with respect to the chemical potentials. 
In the saddle point approximation to this integral one has to look for the dominant values of the chemical potentials.

The simplest interpretation of the black hole entropy is that the degeneracy of states should have 
an exponential growth as a function of the charges as the charges become large. 
Presently  we shall show that this is indeed the case. 
The degeneracy of states can be calculated as the inverse Laplace transform of the index~$\CI$
with respect to the chemical potentials. 
We focus on asymptotic formulas in the large-charge \emph{Cardy-like} limit, 
in which this integral is dominated by very small values of chemical potentials $\sigma,\tau$ for angular momenta---plus possibly 
an integer, according to the discussion above. 
 We perform our calculations in two different ways, one of which   is exact in~$N$   and the other   uses the large-$N$ limit.
In both methods our starting point is the matrix integral representation~\eqref{index_other_writing} of the index.
In the first method, valid for finite~$N$, we first take the Cardy-like limit and then calculate the saddle points. 
In the second method we look directly for large-$N$ saddle points of the index using the matrix integral 
representation~\eqref{index_other_writing}. 
Then we estimate the integral in the large-charge 
limit and show that the saddle point indeed reproduces the black hole entropy for~$n_0=-1$. 

In the case of ${\cal N}=4$ SYM theory,  the large-$N$ limit of the index defined in \cite{Kinney:2005ej} 
has been analyzed in~\cite{Benini:2018ywd} using a different Bethe-ansatz-type 
representation, in which the solutions of the Bethe-ansatz equations play the role of saddle-points. One  
of these solutions leads to the black hole entropy and, although this has a complex value of gauge 
holonomies $u_i-u_j$  to begin with, it reduces to~$u_i-u_j=0$ in the Cardy-like limit, consistent with our saddle-point analysis. 
The Cardy-like limit has been analyzed in~\cite{Choi:2018hmj, Choi:2018vbz,Honda:2019cio,ArabiArdehali:2019tdm} for  
four-dimensional~$\CN=4$ SYM, and some extensions have also been considered \cite{Honda:2019cio}.
Some of these analyses assume a saddle point in the relevant range and show that it leads to the correct black hole 
entropy, while others use an auxiliary matrix model or additional flavor chemical potentials
in order to find the saddle points. 
In this paper we avoid these assumptions: we study the index~\eqref{index_other_writing} 
for a large class of~$\CN=1$ SCFTs 
in which we directly find the saddle points in the two methods mentioned above, and show that the value at the leading saddle leads to the black hole entropy. 
In the process there are some mathematical subtleties that we point out and address.

Our main result is that the index of ${\cal N}=1$ SCFTs in the Cardy-like limit has a universal saddle controlled by the anomalies $a,c$, which describes an asymptotic growth of states accounting for the black hole entropy. We further prove that for a large class of theories this is the dominant saddle. 
For these theories, the index at leading and first subleading order in $\t, \s \to 0$ reads
\be
\log \CI(\t,\s; n_0) \,=\,   2\pi \i \, \dfrac{3\tau+3\sigma\pm1}{27\tau\sigma} \, (3c-5a) 
+ 2\pi \i\,\dfrac{\tau+\sigma \pm 1}{3\tau\sigma}\,(a-c) + \mathcal{O}(1)
\,,  \qquad  n_0\=\mp1 \,,
\label{ourmain}
\ee
which is valid \emph{at finite $N$}. 
For $n_0=0$, on the other hand, the asymptotic growth is controlled by a universal term proportional to $\frac{\sigma+\tau}{\sigma\tau}(a-c)$  \cite{DiPietro:2014bca} (which in some special instances can receive corrections \cite{Ardehali:2015bla}). 
Using the constraint~\eqref{ConstraintFT} we can rewrite the result above as 
\be
\log \CI(\t,\s;n_0) \=   
\frac{2\varphi^3}{27 \pi^2\t \s} \, (5a-3c) + \frac{2\varphi}{3 \t\s}\,(a-c) + \mathcal{O}(1)\,.
\label{pippo}
\ee
Thus we find that, at least for holographic theories at large~$N$, $\log \CI$ agrees precisely with the Casimir energy-like prefactor ${\cal F}$, that we computed in~\cite{Cabo-Bizet:2018ehj}. 
The formulas~(\ref{ZFIsplit}) and~(\ref{ourmain})
are related  similarly to the analogous formulas in 2d CFTs, where the 
vacuum energy controls the growth of states of the black hole via the Cardy formula as e.g.~in~\cite{Strominger:1996sh}. 
This observation suggests the existence of a modular-like symmetry in these 4d observables that would very 
interesting to explore further.  Observations of a similar nature have been made in~\cite{Shaghoulian:2016gol}.

In our Cardy-like limit the charges and the saddle point values of the chemical potentials behave as follows. 
If~$\upsilon$ is a scale parameter such that $\sigma,\tau=\mathcal{O}(\upsilon)$ as $\upsilon\to0$, 
then the charges scale as
\bea
Q \= \upsilon^{-2} q\,, \quad J_{1,2} \= \upsilon^{-3} j_{1,2} \,,
\eea
in order to preserve the non-linear constraint that they obey in the extremization process~\cite{Cabo-Bizet:2018ehj}. 
Correspondingly the saddle-point values of the angular velocities~$\t, \s $ scale as $\frac{Q}{J_{1,2}} + O\bigl(\upsilon^2 \bigr)$
at~$n_0=- 1$.
At large~$N$ the entropy scales as 
\bea
S \= \pi \Bigl(\sqrt{3} |Q| - 4 c \, \frac{J_1+J_2}{\sqrt{3} |Q|} \Bigr) + \mathcal{O}(1) \,,
\eea
as consistent with the Bekenstein-Hawking entropy of AdS$_5$ black holes given by~\cite{Kim:2006he}  
\be 
S_{\text{BH}}  \= \pi \sqrt{3 Q^2 - 8 c \bigl(J_1+J_2\bigr)} \, .
\ee  
Our formula~\eqref{ourmain} for the index also contains information about the first subleading corrections in~$\frac{1}{N}$. 
For ${\cal N}=4$ SYM this observation was made in \cite{Honda:2019cio}.

The plan of the paper is as follows. In Section~\ref{sec:Cardy} we consider  the Cardy-like  limit of the index 
for~${\cal N}=1$ field theories with a Lagrangian description. 
Firstly, we show that in a  non-empty neighborhood  including zero, there exists a universal saddle point 
corresponding to all vanishing gauge holonomies, which gives rise to the asymptotic growth (\ref{pippo}), reproducing the entropy of the AdS$_5$ black holes.
Since this is controlled by anomalies, the result could extend for non-Lagrangian theories as well.
Secondly, we study the matrix model in its complete domain of definition for generic superconformal
quiver gauge theories, and prove the uniqueness of the universal saddle point in large classes of  ${\cal N}=1$ SCFTs.
In Section~\ref{Sec:largeN} we begin a study of the large-$N$ limit of the index. We find a family of saddle points
at real values of gauge holonomies, of which the dominant saddle in the large-$N$ limit is the one where
all the holonomies clump up at the point where they all vanish. 
We show that the index in the Cardy-like limit at this saddle point leads to the black hole entropy at large~$N$.

\vspace{0.2cm}

\ndt {\bf Note added:} While this paper was being completed we received~\cite{Kim:2019yrz} 
which also presents results for~$\CN=1$ 4d SCFTs. These results have some overlap with Section 2 of this paper.

\section{The Cardy-like limit of the superconformal index \label{sec:Cardy}}

The Cardy-like (``high-temperature'') limit of the superconformal index was studied in \cite{DiPietro:2014bca,DiPietro:2016ond} using effective field theory arguments on $S^3\times S^1$ backgrounds, and in \cite{Ardehali:2015bla} using asymptotic properties of the elliptic gamma functions.\footnote{See also \cite{Ardehali:2015hya,Hosseini:2016cyf,Hwang:2018riu} for related work.} Assuming real values of the fugacities and taking the limit of small chemical potentials $\sigma,\tau$, these works found a universal exponential contribution to the index weighted by the 't Hooft anomaly ${\rm Tr}R \propto (a-c)$ (this can receive corrections in  special instances \cite{Ardehali:2015bla}). It follows that in this limit the asymptotic growth of the index at large $N$ is not enough to reproduce the Bekenstein-Hawking entropy of holographically dual AdS$_5$ black holes.
More recently, \cite{Choi:2018hmj,Honda:2019cio,ArabiArdehali:2019tdm} considered the $\mathcal{N}=4$ SYM index and showed that for suitable {\it complex} values of the flavor fugacities the Cardy-like limit yields an exponential $\mathcal{O}(N^2)$ growth at large $N$, which reproduces the entropy function of~\cite{Hosseini:2017mds}.

In this section we enhance these results and study the Cardy-like limit of general $\mathcal{N}=1$ superconformal field theories, without having to introduce any flavor fugacities.
In the notation introduced in Section~\ref{sec:Intro}, we set $n_0=\pm 1$ and take the limit $\sigma, \tau\to 0$. We emphasize again that this is a different limit than the one obtained by setting $n_0=0$ and $\sigma, \tau\to 0$. We prove the existence of a universal saddle point in the index and, with some additional assumptions, we check that this dominates the index in a very large class of theories.
The saddle point value is controlled by both the ${\rm Tr}R^3$ and the ${\rm Tr}R$ 't Hooft anomalies. In the large $N$ limit this yields the entropy function which upon Legendre transform reproduces the Bekenstein-Hawking entropy of the supersymmetric AdS$_5$ black holes of~\cite{Gutowski:2004ez,Chong:2005hr}. These black holes are solutions to five-dimensional minimal gauged supergravity and their properties are indeed expected to be described by any four-dimensional $\mathcal{N}=1$ SCFT with a weakly-coupled gravity dual.

\subsection{Derivation of the saddle point}\label{asymptotics_general}

We consider an $\mathcal{N}=1$ field theory with gauge group $G$ and a non-anomalous $U(1)_R$ R-symmetry. 
When the theory flows to a superconformal fixed point, we pick the R-symmetry that enters in the IR superconformal algebra. 
%Although conformality is not really necessary to define the index, we assume that the theory flows to an IR fixed point and we pick the R-symmetry that enters in the IR superconformal algebra. 
 A class of~$\CN=1$ superconformal field theories was first proposed and its finiteness checked in perturbation theory in \cite{Parkes:1984dh,Parkes:1985hj,West:1984dg}.
 We label by the letter $I$ the chiral multiplets of the theory; these sit in the representation ${\rm R}_I$ of the gauge group and have R-charge $r_I$. 
 Throughout the paper we assume $0<r_I<2$ for all chiral multiplets. 

The index discussed in the introduction has the integral representation
\be\label{index_other_writing}
\mathcal I(\sigma,\tau;n_0) 
= \frac{1}{|\mathcal{W}|}\int \prod_{i=1}^{\text{rk}(G)} \diff u_i   \, \mathcal{Z}_{\rm vec} \mathcal{Z}_{\rm chi}\, \ ,
\ee
where the integration variables~$u_i$, $i=1,\ldots,{\rm rk}(G)$, parameterize the maximal torus of $G$ and run over the real range~$(-\frac12,\frac12 ]$, 
 $|\mathcal{W}|$ is the order of the Weyl group of $G$, and $\mathcal{Z}_{\rm vec}$, $\mathcal{Z}_{\rm chi}$ denote the vector multiplet and chiral multiplet contribution, respectively. These read:
\begin{align}\label{ZvecZchi}
\mathcal{Z}_{\rm vec} &= (\rme^{2\pi \ii\sigma};\rme^{2\pi \ii\sigma})^{{\rm rk}(G)}(\rme^{2\pi \ii\tau};\rme^{2\pi \ii\tau})^{{\rm rk}(G)} \prod_{ \alpha \in \Delta_+}   \Ge(\alpha\cdot u + \sigma + \tau ; \sigma , \tau)  \Ge( - \alpha\cdot u + \sigma + \tau ; \sigma , \tau) \, ,\nn\\[1mm]
\mathcal{Z}_{\rm chi} &= \prod_{I\in {\rm chirals}} \prod_{ \rho \in \text{R}_I}\Ge (z_I ; \sigma,\tau ) \ ,
\end{align}
where $\Delta_+$ is the set of positive roots, denoted by $\alpha$, of the gauge group $G$, and $\rho$ is the weight of the gauge representation $R_I$.  The complex parameters $\sigma,\tau$ are taken in the upper half-plane,
\be
{\rm Im}\,\sigma > 0\ ,\quad {\rm Im}\,\tau >0\ .
\ee
The variable $z_I$ appearing in the chiral multiplet contribution is defined as
\be \label{defzI}
z_I  =  \rho\cdot u+\frac{r_I}{2} (\sigma+\tau-n_0) \,, \quad \text{ with } \rho\in \text{R}_I \,.
\ee
Notice the dependence on the choice of integer $n_0$, which will play a central role in the following. 
The Pochhammer symbol, encoding the contribution of the vanishing roots of $G$, is defined for $w,q \in \mathbb{C}$, with $|q|<1$,  as 
\be
(w;q)= \prod_{j=0}^\infty (1-wq^j)\ .
\ee
Finally, the elliptic gamma function is defined as
\be\label{GammaeDef}
\Ge(z;\sigma,\tau)\, = \, 
\prod_{j,k=0}^{\infty}
\frac{1-\rme^{-2 \pi \ii z} \, \rme^{2 \pi \ii \sigma (j+1)} \, \rme^{2 \pi \ii \tau(k+1)}}{1-\rme^{2 \pi \ii z} \, 
\rme^{2 \pi \ii \sigma j} \, \rme^{2 \pi \ii \tau k}} \,.
\ee
 This is a meromorphic function in $z\in \mathbb{C}$, with simple poles at $z= -j\sigma -k\tau +l$ and simple zeros at $z=(j+1)\sigma+(k+1)\tau +l$, where $j,k\in \mathbb{Z}^{\geq 0}$ and $l \in \mathbb{Z}$. 
A discussion of its properties can be found in \cite{Felder,Spiridonov:2010em,Spiridonov:2012ww}.\footnote{The vector multiplet contribution to the index can also be expressed in terms of the Jacobi theta function, which is defined as
$$
\theta_0(z;\tau) = (w;q)(q/w;q)\ ,
$$
with $w=\rme^{2\pi \ii z}$ and $q=\rme^{2\pi \ii\tau}$.
Using Proposition~3.2 in \cite{Felder}, one can see that the contribution of the non-vanishing roots can be written as  $$\Ge(\alpha\cdot u + \sigma + \tau ; \sigma , \tau)  \Ge( - \alpha\cdot u + \sigma + \tau ; \sigma , \tau)  = \frac{1}{\Ge(-\alpha\cdot u;\sigma,\tau)\Ge(\alpha\cdot u;\sigma,\tau)}=\theta_0(\alpha\cdot u;\sigma)\theta_0(- \alpha\cdot u;\tau)\ .$$
These alternative expressions are often used in the literature on the superconformal index.}

The integral representation \eqref{index_other_writing} was derived in \cite{Cabo-Bizet:2018ehj} as the partition function of supersymmetric field theories on $S^3\times S^1$, with complexified angular chemical potentials $\sigma,\tau$ for rotation in $S^3$, and holonomy for the background R-symmetry gauge field $A$ being given by $\int_{S^1} A = \ii \varphi$, with $\varphi=\pi \ii(\sigma+\tau -n_0)$.
It is a slight modification of the familiar integral representation of the superconformal index, which corresponds to $n_0=0$.

We want to study a Cardy-like limit of \eqref{index_other_writing} where the angular chemical potentials $\sigma,\tau$ become small. We stress that in this limit the holonomy $\varphi$ for the R-symmetry gauge field is kept finite if $n_0=\pm1$. We will perform the analysis for $n_0=-1$ (this can be straightforwardly adapted to the case $n_0=+1$). Our main technical tool will be a uniform estimate for the elliptic gamma function given in Proposition 2.12 of \cite{Rains:2006dfy} and conveniently adapted for applications to the superconformal index in \cite{Ardehali:2015bla} whose notations we follow.
Let us define $\sigma = \upsilon \check{\sigma}$, $\tau = \upsilon \check{\tau}$ and consider the limit $\upsilon \to 0^+$ with $\check{\sigma},\check{\tau}$ finite.
Then outside of an $\mathcal{O}(\upsilon)$ neighborhood of the zeros and of the poles, the elliptic gamma function is uniformly estimated by
\be\label{estimate_Rains}
\log \Ge\left( x + \frac{r}{2}(\sigma + \tau); \sigma,\tau \right) = 2\pi i \left[ -\frac{\kappa(x)}{12\sigma\tau} + (r-1) \frac{\sigma+\tau}{4\sigma\tau}\left( \vartheta(x)  - \frac{1}{6} \right)  \right] + \mathcal{O}(1)\ .
\ee
Here and in the following, by $\mathcal{O}(1)$ we denote subleading terms in the small $\upsilon$ expansion  (this also includes possible logarithmically divergent terms). Moreover we have defined the functions
\be
\kappa(x) = \{x\} (1-\{x\})(1-2\{x\})  \ , \qquad \vartheta(x) = \{x\}(1-\{x\})\ ,
\ee
where $\{x\}=x - \lfloor x\rfloor$ is the fractional part of $x$. These functions are periodic with period~1 and satisfy
\be
\kappa(-x)=-\kappa(x)\ ,\qquad  \vartheta(-x)=\vartheta(x)\ .
\ee

We now estimate the index \eqref{index_other_writing} in the limit of small angular chemical potentials defined above. We start from the chiral multiplet contribution $\mathcal{Z}_{\rm chi}$  in \eqref{ZvecZchi}, which depends on the variable $z_I$ in \eqref{defzI}. In order to apply the estimate above 
we thus need to set the variable $x$ in \eqref{estimate_Rains} to
\be
x = \rho\cdot u - \frac{r n_0 }{2}\ ,\qquad\text{with}\ n_0=-1\ .
\ee
In this way we obtain
\be
\log \mathcal{Z}_{\rm chi} = 2\pi \ii \sum_{I\in {\rm chiral}}\sum_{\rho\in {\rm R}_I} \left[ -\frac{\kappa(\rho\cdot u + \frac{r_I}{2})}{12\sigma\tau} + (r_I-1) \frac{\sigma+\tau}{4\sigma\tau}\left( \vartheta(\rho\cdot u + \tfrac{r_I}{2})  - \frac{1}{6} \right)  \right] + \mathcal{O}(1)\,,
\ee
which is uniform over all values of $\rho\cdot u + \frac{r_I}{2}$ since we are assuming $0<r_I<2$, and we are thus staying away from the zeros and poles of the elliptic gamma functions in $\mathcal{Z}_{\rm chi}$.

As for the vector multiplet contribution, the estimate \eqref{estimate_Rains} leads to \cite{Ardehali:2015bla}:
\be\label{estimate_vectors}
\log \left[\Ge(\alpha\cdot u + \sigma + \tau ; \sigma , \tau)  \Ge( - \alpha\cdot u + \sigma + \tau ; \sigma , \tau) \right] =  \pi \ii  \frac{\sigma+\tau}{\sigma\tau}\left( \vartheta(\alpha\cdot u)  - \frac{1}{6} \right)  + \mathcal{O}(1)\ .
\ee
We recall that a priori this  is valid outside neighborhoods of size $\mathcal{O}(\upsilon)$  around the points $\alpha\cdot u\in \mathbb{Z}$, where the gamma functions in \eqref{estimate_vectors} vanish. One should therefore ask whether it is legitimate to use the estimate in the whole range of the gauge holonomies to approximate the integrand (especially because we will find later that the saddle point of the estimate lies in $\alpha\cdot u=0$). A related issue is whether this estimate for the integrand gives an equally good estimate of the integral. It has been argued in \cite{Ardehali:2015bla} that this is the case, as such subtleties give rise to negligible errors at the order of precision considered; we expect the same is true in our setup. 
 Also including the contribution from the asymptotics of the Pochhammer symbols, the full vector multiplet contribution $\mathcal{Z}_{\rm vec}$ reads
\be
\log \mathcal{Z}_{\rm vec} = -\pi \ii\,  \frac{\sigma+\tau}{12\sigma\tau}{\rm dim}\, G +  \pi \ii \, \frac{\sigma+\tau}{\sigma\tau}\sum_{\alpha\in\Delta_+} \vartheta(\alpha\cdot u)    + \mathcal{O}(1)\ .
\ee 

Putting everything together, we arrive at the following estimate for the full integrand
\be\label{log_integrand}
\log   \left( \mathcal{Z}_{\rm vec} \mathcal{Z}_{\rm chi} \right)  = -\pi \ii\,\frac{\sigma+\tau}{12\sigma\tau}\,{\rm Tr}R +  \pi \ii \, \frac{\sigma+\tau}{2\sigma\tau} V_1 + \frac{\pi \ii}{6\sigma\tau} V_2 + \mathcal{O}(1)\,,
\ee
where
\begin{align}\label{V2andV1}
V_2 &= -\sum_{I\in {\rm chirals}}\sum_{\rho_I\in {\rm R}_I} \kappa(\rho_I\cdot u + \tfrac{r_I}{2})\ ,\nn\\[1mm] 
V_1  &= 2\sum_{\alpha\in\Delta_+} \vartheta(\alpha\cdot u)  + \sum_{I\in {\rm chirals}}\sum_{\rho_I\in {\rm R}_I}(r_I-1)\vartheta(\rho_I\cdot u + \tfrac{r_I}{2})\ .
\end{align}

The first term on the right hand side of \eqref{log_integrand}, proportional to the 't Hooft anomaly ${\rm Tr}R$ (see \eqref{tHooftanomalies} below for its definition) and independent of the gauge holonomies, was first obtained in \cite{DiPietro:2014bca}. The remaining two terms determine an effective potential for the gauge holonomies and are similar to those first found in \cite{Ardehali:2015bla} (see also  \cite{DiPietro:2016ond}). However we have two important differences: the argument of the functions $V_1$, $V_2$ is $\rho\cdot u + \frac{r}{2}$ instead of $\rho\cdot u$ (because we are taking $n_0=-1$ while the setup of \cite{Ardehali:2015bla,DiPietro:2016ond} corresponds to $n_0=0$), and $\sigma,\tau$ are generically complex (while they were assumed purely imaginary in \cite{Ardehali:2015bla,DiPietro:2016ond}, which corresponds to real fugacities $p=\rme^{2\pi \ii \sigma}$, $q=\rme^{2\pi \ii \tau}$). 
 Recently, similar asymptotic formulae have been discussed in the specific case of $\mathcal{N}=4$ SYM in \cite{Honda:2019cio,ArabiArdehali:2019tdm} by allowing for complex values of the fugacities. In this case, two chemical potentials for the flavor symmetries play a role analogous to our discrete holonomy $n_0 = \pm 1$ for the R-symmetry; we will comment more on this in Section~\ref{SYMsection}.

In a general theory where the matter multiplets sit in different representations of the gauge group and have different R-charges the effective potential is quite complicated, as it depends on $\{\rho\cdot u + \frac{r_I}{2}\} = \rho\cdot u + \frac{r_I}{2} - \lfloor  \rho\cdot u + \frac{r_I}{2}\rfloor$, and the integer part $\lfloor\rho\cdot u + \frac{r_I}{2}\rfloor$ depends on the value of the gauge holonomies as well as on the weight vector $\rho$ and the R-charge $r_I$. 
However a central point of our analysis is that we can say something universal when the gauge holonomies are sufficiently close to zero. In order to make this precise, let us consider the intersection of all the intervals such that the inequality
 \be\label{nice_range}
 -\frac{r_I}{2}\leq \rho_I \cdot u< 1-\frac{r_I}{2}
 \ee 
is satisfied for all chiral multiplets and all weight vectors, so that $\{ \rho\cdot u + \frac{r_I}{2} \} = \rho\cdot u + \frac{r_I}{2} $ there.\footnote{In the case of a theory with non-chiral matter content, for any contribution from a weight vector $\rho$ we have an equivalent contribution from the weight vector $-\rho$. In this case the inequality \eqref{nice_range} can be written as $|\rho_I\cdot u| \leq  {\rm min}(\frac{r_I}{2},1-\frac{r_I}{2})$.}  Note that this region is not empty as it contains a neighborhood of $u_i = 0$, $i=1,\ldots,{\rm rk}(G)$. % (recall that we assumed $0<r<2$ for all chiral multiplets).  
In this region the term in the effective potential controlling the leading divergence $\frac{1}{\sigma\tau}$ reads
\begin{align}
V_2 \,&=\, -\sum_{I\in {\rm chirals}}\sum_{\rho_I\in {\rm R}_I} (\rho_I\cdot u + \tfrac{r_I}{2})  (1-\rho_I\cdot u - \tfrac{r_I}{2})(1-2\rho_I\cdot u - r_I) \nn\\[1mm]
 \,&=\, \frac{{\rm Tr}R - {\rm Tr} R^3}{4}-3  \sum_{I\in {\rm chirals}}\sum_{\rho_I\in {\rm R}_I}(r_I-1)(\rho_I\cdot u)^2  \nn\\[1mm]
 \,&=\, \frac{{\rm Tr}R - {\rm Tr} R^3}{4} +6   \sum_{\alpha\in \Delta_+}(\alpha\cdot u)^2\ ,
\label{eyecontact}
\end{align}
where we have introduced the 't Hooft anomalies
\begin{align}\label{tHooftanomalies}
{\rm Tr}R &= {\rm dim} G + \sum_{I\in \text{chirals}}{\rm dim}{\rm R}_I \,(r_I-1)\ , \nn\\
{\rm Tr}R^3 &= {\rm dim} G + \sum_{I\in \text{chirals}}{\rm dim}{\rm R}_I\, (r_I-1)^3\ .
\end{align}
Moreover the second line has been simplified by using the following consequences of anomaly cancellation:
\begin{align}
\sum_{I\in {\rm chirals}}\sum_{\rho_I\in {\rm R}_I} (\rho_I \cdot u)^3  = 0\ ,\nn\\[1mm]
\sum_{I\in {\rm chirals}}\sum_{\rho_I\in {\rm R}_I} \rho_I \cdot u  = 0\ ,\nn\\[1mm]
\sum_{I\in {\rm chirals}}\sum_{\rho_I\in {\rm R}_I}(r_I-1)^2\,\rho_I \cdot u = 0\ ,\label{anomaly_canc_1}
 \end{align}
while in the last line we used
\be\label{anomaly_canc_2}
2\sum_{\alpha\in \Delta_+} (\alpha\cdot u)^2  + \sum_{I\in {\rm chirals}}\sum_{\rho_I\in {\rm R}_I}(r_I-1)(\rho_I\cdot u)^2  = 0\ .
\ee
Eqs.~\eqref{anomaly_canc_1} follow from cancellation of the gauge $G^3$ anomaly, the mixed gauge--gravitational anomaly and the mixed $G - U(1)_R^2$ anomaly, respectively. Eq.\ \eqref{anomaly_canc_2} follows from cancellation of the $U(1)_R - G^2$ anomaly, which corresponds to the requirement that the R-symmetry is preserved at the quantum level.

We can also evaluate the term $V_1$ in the range \eqref{nice_range} of the gauge holonomies. We find:
\begin{align}
V_1  \,&=\, 2\sum_{\alpha\in\Delta_+} \vartheta(\alpha\cdot u)  + \sum_{I\in {\rm chirals}}\sum_{\rho_I\in {\rm R}_I}(r_I-1)(\rho_I\cdot u + \tfrac{r_I}{2})(1-\rho_I\cdot u - \tfrac{r_I}{2}) \nn\\[1mm]
\,&=\,  \frac{{\rm Tr}R - {\rm Tr} R^3}{4} +  2\sum_{\alpha\in\Delta_+} \left(\vartheta(\alpha\cdot u) + (\alpha\cdot u)^2 \right)  \ ,
\end{align}
where again we exploited anomaly cancellation by using the third in \eqref{anomaly_canc_1} as well as \eqref{anomaly_canc_2}. As an aside, we observe that when the gauge group is a product of $U(N)$ or $SU(N)$ factors, $\alpha\cdot u \in (-1,1)$ and in this range we have $\vartheta(\alpha\cdot u) = |\alpha\cdot u| - (\alpha\cdot u)^2$. It follows that for these gauge groups, $V_1$ can be written more simply as
\be
V_1  \,=\, \frac{{\rm Tr}R - {\rm Tr} R^3}{4} +  2\sum_{\alpha\in\Delta_+} |\alpha\cdot u|  \ .
\ee

Plugging the expressions for $V_2$ and $V_1$ in \eqref{log_integrand}, we conclude that in the region \eqref{nice_range}, the integrand of the index \eqref{index_other_writing} is approximated by
\begin{align}
\log   \left( \mathcal{Z}_{\rm vec} \mathcal{Z}_{\rm chi} \right) &= -  \pi \ii \, \frac{3\sigma+3\tau+1}{24\sigma\tau}  {\rm Tr} R^3 + \pi \ii\,\frac{\sigma+\tau+1}{24\sigma\tau}\,{\rm Tr}R \nn\\[1mm]
&\quad+ \frac{\pi \ii}{\sigma\tau} \, \sum_{\alpha\in\Delta_+}\left[ (\alpha\cdot u)^2 +  (\sigma+\tau) \left(\vartheta(\alpha\cdot u) + (\alpha\cdot u)^2 \right)  \right] + \mathcal{O}(1)\ .
\end{align}

Recall that $V_2$ controls the most divergent contribution to the effective potential in the Cardy-like limit. It is clear that this is extremized in $\alpha\cdot u = 0$ for all roots, which is achieved by taking simply $u_i=u={\rm const}$ for all $i$. The same configuration also extremizes $V_1$, which also is an even function of $\alpha\cdot u$.

From \eqref{log_integrand} we see that if 
\be\label{conditions_dominate}
{\rm Re} \left( \frac{\ii}{\sigma\tau} \right) < 0 \qquad \text{and}\qquad {\rm Tr}R^3 - {\rm Tr} R >0\ ,
\ee
then this extremum  dominates the Cardy-like limit of the integrand in the region \eqref{nice_range}.
For $\mathcal{N}=1$ superconformal theories, the 't Hooft anomalies ${\rm Tr}R^3$ and ${\rm Tr}R$ can be traded for the $a$ and $c$ central charges appearing in the Weyl anomaly via the linear relations
\be\label{ac_anomalies}
a = \frac{3}{32}\left( 3{\rm Tr}R^3 - {\rm Tr}R \right)\ ,\qquad
c= \frac{1}{32}\left( 9{\rm Tr}R^3 - 5{\rm Tr}R \right) \ ,
\ee
so that the second condition in \eqref{conditions_dominate} is equivalent to 
\be 
\label{HMbound}
3c-2a>0\ ,\ee 
which is  indeed satisfied very generally in SCFTs \cite{Hofman:2008ar}.
Assuming\footnote{We will later show that this assumption is correct for very large classes of theories.} that there are no competing minima of the effective potential outside  the region \eqref{nice_range}, we conclude that for ${\rm Re} \left( \frac{\ii}{\sigma\tau} \right) < 0$ the Cardy-like limit of the index \eqref{index_other_writing} is
\be\label{result_Cardy_limit}
\log\mathcal I(\sigma,\tau;n_0=-1) = -  \pi \ii \, \frac{3\sigma+3\tau+1}{24\sigma\tau} \, {\rm Tr} R^3 + \pi \ii\,\frac{\sigma+\tau+1}{24\sigma\tau}\,{\rm Tr}R + \mathcal{O}(1)\,.
\ee

Using \eqref{ac_anomalies}, the result \eqref{result_Cardy_limit} can be written as
\be\label{result_integrand_divergent}
\log\mathcal I(\sigma,\tau;n_0=-1) = -  2\pi \ii \, \frac{3\sigma+3\tau+1}{27\sigma\tau} \, (5a-3c) + 2\pi \ii\,\frac{\sigma+\tau+1}{3\sigma\tau}\,(a-c) + \mathcal{O}(1)\ ,
\ee
which can also be rearranged as
\be
\log\mathcal I(\sigma,\tau;n_0=-1) = -  4\pi \ii \, \frac{3\sigma+3\tau+1}{27\sigma\tau} \, (3c-2a) - 4\pi \ii\,\frac{\sigma+\tau}{3\sigma\tau}\,(a-c) + \mathcal{O}(1)\ .
\ee

In terms of the variables $\omega_1 = 2\pi \ii \sigma$, $\omega_2=2\pi \ii\tau$ and $\varphi = \frac{1}{2}(\omega_1+\omega_2 - 2\pi \ii n_0)$ 
used in~\cite{Cabo-Bizet:2018ehj}, this may be written as:
\be\label{result_Cardy_limit_old_var}
\log\mathcal I(\sigma,\tau;n_0) = -  \, \frac{8\varphi^3}{27\omega_1\omega_2} \, (5a-3c) - \frac{8\pi^2\varphi}{3\omega_1\omega_2}\,(a-c) + \mathcal{O}(1)\,,
\ee
where we emphasize that only the quadratically and linearly divergent terms in the small $\omega_1,\omega_2$ limit have been determined by our method, and we have taken $n_0=-1$.

For holographic theories at large $N$, that is after taking $a=c$, this result reproduces the Cardy-like limit of the entropy function \cite{Hosseini:2017mds} which controls the Bekenstein-Hawking entropy of the supersymmetric AdS$_5$ black holes of~\cite{Gutowski:2004ez,Chong:2005hr}. These black holes are constructed within minimal five-dimensional gauged supergravity, hence the corresponding ensemble of states is expected to exist in any SCFT$_4$ with a weakly-coupled holographic dual.
On the gravity side, the same entropy function was derived from a particular BPS limit of black hole thermodynamics in \cite{Cabo-Bizet:2018ehj}. 
Additionally, we observe that the result \eqref{result_Cardy_limit_old_var} for~$\log \CI$ precisely equals the supersymmetric Casimir energy 
prefactor~$\mathcal{F}$ of \cite{Cabo-Bizet:2018ehj}, evaluated at the same order in the Cardy-like limit. 
%The result \eqref{result_Cardy_limit_old_var} for~$\log \CI$ can also be compared with the supersymmetric Casimir energy 
%prefactor~$\mathcal{F}$ of \cite{Cabo-Bizet:2018ehj}, evaluated at the same order in the Cardy-like limit. 
This observation, anticipated in~\cite{Cabo-Bizet:2018ehj}, suggests some modular properties of the supersymmetric partition function 
which would be interesting to investigate. 
%We see that for holographic theories at large $N$ these two functions agree, as anticipated in  \cite{Cabo-Bizet:2018ehj}.

In the regime ${\rm Re}\left(\frac{\ii}{\sigma\tau}\right)>0$, the Cardy-like limit of the integrand \eqref{log_integrand} is dominated by maxima of $V_2$,
such that $V_2>0$ at the maximum.  
The interpretation of this regime for $n_0=-1$ seems not obvious and we postpone it to future work. 
Here we just observe that~$V_2$ contains the generally negative contribution~$\text{Tr} \, R - \text{Tr} \, R^3$,
and a positive value at the maximum may not be easy to achieve. However, in the regime ${\rm Re}\left(\frac{\ii}{\sigma\tau}\right)>0$ the saddle corresponding to the supersymmetric AdS$_5$ black holes of~\cite{Gutowski:2004ez,Chong:2005hr} is found by taking the Cardy-like limit of the $n_0=+1$ index. Indeed starting with $n_0=1$ and repeating the analysis of this section leads to a slightly different form of $V_2$, which is maximized in $\alpha\cdot u=0$ and whose saddle point value is $V_2|_{\alpha\cdot u=0}=\frac{1}{4}({\rm Tr}R^3 - {\rm Tr} R)>0$. As a result, whenever there are no competing saddles the Cardy-like limit of the $n_0=1$ index is 
\be
\log\mathcal I(\sigma,\tau;n_0=1) = -  \pi \ii \, \frac{3\sigma+3\tau-1}{24\sigma\tau} \, {\rm Tr} R^3 + \pi \ii\,\frac{\sigma+\tau-1}{24\sigma\tau}\,{\rm Tr}R + \mathcal{O}(1)\,.
\ee
Note that in the variables $\omega_1,\omega_2,\varphi$ this again takes the form \eqref{result_Cardy_limit_old_var}.

Next we analyse the effective potential beyond the region~\eqref{nice_range} for certain classes of ${\cal N}=1$ superconformal gauge theories with known holographic duals. For these examples, we will show that the extremum at the origin discussed above does dominate the index.

\subsection{$\mathcal{N}=4$ SYM with gauge group $G = SU(N)$}\label{SYMsection}

As a first illustrative example, we consider $\mathcal{N}=4$ SYM with gauge group $G = SU(N)$,
which here we shall view as a special ${\cal N}=1$ theory. 
We take gauge holonomies $u_i \in (-\frac{1}{2},\frac{1}{2}]$, $i=1,\ldots,N$, subject to the constraint $\sum_i^N u_i = 0$. In addition to the $SU(N)$ vector multiplet, we have three chiral multiplets, also in the adjoint representation, with R-charge $r =2/3$. The adjoint representation has roots $\alpha_{ij}$, such that $\alpha_{ij} \cdot u = u_i - u_j \equiv u_{ij}$. The assumed range of the gauge holonomies implies $u_{ij} \in (-1,1)$.

We discuss the case $n_0=-1$. The functions $V_2$ and $V_1$ defined in \eqref{V2andV1} read:
\begin{align}
\label{V2N4}
V_2 &= -3\sum_{ i<j}^N \left[ \kappa(u_{ij} + \tfrac{1}{3}) + \kappa(-u_{ij} + \tfrac{1}{3})    \right]   -3 (N-1)\,\kappa(\tfrac{1}{3}) \ ,\nn\\[1mm]
V_1  &= \sum_{i<j}^N\left[2\vartheta(u_{ij}) -\vartheta(u_{ij} + \tfrac{1}{3}) - \vartheta( -u_{ij} + \tfrac{1}{3})   \right]   - (N-1) \vartheta(\tfrac{1}{3})\ .
\end{align}

\begin{figure}
	\centering
\includegraphics[width=7.5cm]{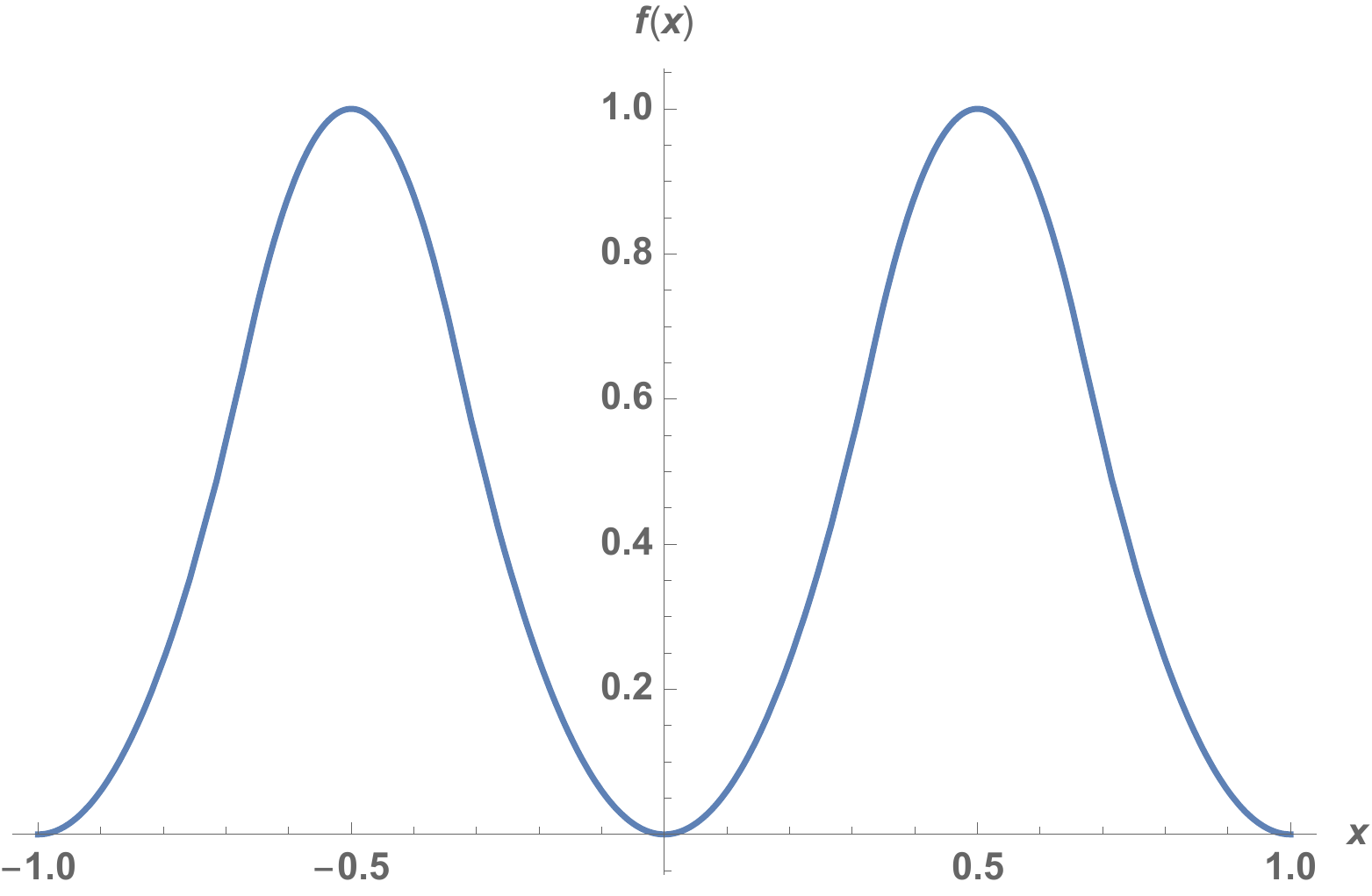}
\caption{The function $f$, determining the leading behavior of the effective potential.\label{functionf}}
\end{figure}

We analyse the function $V_2$, as it controls the leading term of the effective potential in the Cardy-like limit. This can be written as:
\be
V_2 =     \sum_{ i<j}^N f(u_{ij})  - \frac{2}{9} (N^2-1)   \ ,
\ee
where the function
\be
f(x) = 
\begin{cases}
& 6x^2 \qquad \qquad\qquad\qquad \ \text{for } |x| \leq\frac{1}{3} \\
& -12x^2 +12|x| -2      \qquad \text{for }  \frac{1}{3}  < |x| \leq \frac{2}{3}   \\
& 6\, (1-|x|)^2 \qquad\qquad\quad\, \text{for }  \frac{2}{3}  < |x| < 1 \\
\end{cases}
\label{potforsym}
\ee
 is displayed in Figure~\ref{functionf}. It has maxima $f=1$ in $x=\pm 1/2$ and a global
  minimum  $f=0$ in $x=0$ (the points~$x= \pm 1$ are not reached by the chosen range of the gauge holonomies).
If ${\rm Re}\left(\frac{\ii}{\sigma\tau}\right)<0$, then we need to minimize $V_2$. Clearly this is achieved by $u_{ij}=0$, which corresponds to taking $u_i=0$ for all the gauge holonomies. Inserting 
\be
V_2|_{u_i = 0} =V_1|_{u_i = 0} = -\frac{2}{9}(N^2-1) 
\ee 
 into \eqref{log_integrand}, and recalling that for $SU(N)$ $\mathcal{N}=4$ SYM the central charges satisfy $a=c=\frac{N^2-1}{4}$, we arrive at
\be
\log \mathcal{I} =   -4\pi i \, \frac{3\sigma+3\tau+1}{27\sigma\tau} \,c + \mathcal{O}(1)\,,
\ee
in agreement with the general result \eqref{result_integrand_divergent}.
Since we have proven that this is the only minimum of $V_2$, we can conclude that it dominates the index.

Before considering more general classes of theories, we would like to make contact with related approaches to the Cardy-like limit of $\mathcal{N}=4$ SYM taken in \cite{Honda:2019cio,ArabiArdehali:2019tdm} (the comment also applies to \cite{Benini:2018ywd}). By comparing the analyses, it is apparent that the role played by the flavor chemical potentials in \cite{Honda:2019cio,ArabiArdehali:2019tdm} is related to the holonomy for the background $U(1)_R$ gauge field controlled by the integer $n_0$. For definiteness we refer to \cite{Honda:2019cio}, where the flavor chemical potentials were denoted by $m_1,m_2$. 
We find that the shift by $-\frac{n_0r}{2}=\frac{1}{3}$ in the variables \eqref{defzI} matches the particular choice $m_1=m_2=\frac{1}{3}$ in \cite{Honda:2019cio}.  To see why this is the case, let us consider the superconformal index of $\mathcal{N}=4$ SYM, including chemical potentials $m_1,m_2$ for the flavor symmetries with charge $q_1,q_2$:
\be\label{N=4index}
\mathcal{I}_{\mathcal{N}=4} 
=  \, {\rm Tr}\,  (-1)^F \rme^{-\beta \{\mathcal{Q},\bar{\mathcal{Q}}\}  
+2 \pi \ii \sigma (J_1+\frac{1}{2}Q)+2 \pi \ii \tau (J_2+\frac{1}{2}Q)} \rme^{2\pi\ii (m_1q_1 + m_2q_2)} \ .
\ee
We recall that $Q$ generates the R-symmetry in the $\mathcal{N}=1$ superalgebra generated by the supercharges $\mathcal{Q},\bar{\mathcal{Q}}$. In terms of the  charges $R_1,R_2,R_3$ generating the $U(1)^3\in SO(6)_R$ symmetry, one has $Q = \frac{1}{3}(R_1 + R_2+R_3)$ and $q_{1,2}=\frac{1}{2}(R_{1,2}-R_3)$.
In order to recover the universal superconformal index which only uses the symmetries generated by $J_1,J_2$ and $Q$, one should set $m_1=m_2=0$. In our notation, this corresponds to the $n_0=0$ index. On the other hand, the $n_0=\mp 1$ index is retrieved from \eqref{N=4index} by setting $m_1=m_2=\pm\frac{1}{3}$. Doing so and using $q_1+q_2 = \frac{3}{2}(Q - R_3)$ as well as the relation $\frac{R_3}{2}=- J_1$ (mod 1), one has
\be
\rme^{2\pi\ii (m_1q_1 + m_2q_2)} = \rme^{\pm 2\pi\ii(J_1+\frac{Q}{2})}
\ee
and therefore
\be
\mathcal{I}_{\mathcal{N}=4} 
=  \, {\rm Tr}\,  (-1)^F \rme^{-\beta \{\mathcal{Q},\bar{\mathcal{Q}}\}  
+2 \pi \ii (\sigma\pm1) (J_1+\frac{1}{2}Q)+2 \pi \ii \tau (J_2+\frac{1}{2}Q)} \ ,
\ee
which is indeed the $n_0=\mp1$ index discussed in Section \ref{sec:Intro}.

\subsection{Superconformal ${\cal N}=1$ quiver gauge theories}

In this section we will consider  superconformal ${\cal N}=1$ quiver gauge theories, with gauge group $G = \prod_{a=1}^\nu SU(N_a)$,
with~$N_a=N$ for all~$a$, and chiral fields transforming in 
adjoint or bi-fundamental representations. To study the associated matrix models arising in the Cardy-like limit we will mainly rely on the anomalies cancellation conditions and we will not need to specify the details of the theories. This  was expected because in the computation of  the index using the path integral 
representation   \cite{Assel:2014paa,Cabo-Bizet:2018ehj}  the cancellation of the gauge anomalies is a necessary consistency condition, while no detailed 
information about the superpotential of the theory is required.

In quiver theories the weight vectors $\rho$ are such that  for any bi-fundamental field  $\Phi_{ab}$,
\be
   \rho_{ij}^{\Phi_{ab}} \cdot u \equiv    u_{ij}^{ab} \equiv u_i^a-u_j^b\, ,
\ee
where $a,b$ label the two nodes of the quiver connected by the bi-fundamental  $\Phi_{ab}$, and 
the gauge holonomies  $u_i^a,  u_j^b$ are in  the Cartan subgroups of the two $SU(N)$ factors, respectively. Chiral fields transforming in the adjoint representation of a gauge group, $\Phi_a$, can be incorporated easily, and their associated gauge holonomies are  
$\rho_{ij}^{\Phi_{a}} \cdot u \equiv    u_{ij}^{aa} \equiv  u_{ij}^{a} \equiv u_i^a-u_j^a$.
For these theories the range of the $u_i^a$  implies $u_{ij}^{ab}\in (-1,1)$.  Below we will assume again that $n_0=-1$, while $n_0=+1$  can be treated analogously. 
With these preliminaries  the function $V_2$ defined in~\eqref{V2andV1} and controlling the leading term in the Cardy-like limit reads
\begin{align}
V_2 &= -\sum_{ i,j}^N  \sum_{ ab} \kappa(u_i^a-u_j^b  + \tfrac{r_{ab}}{2})     \, ,
\end{align}
where the sum is over all bi-fundamental fields $\Phi_{ab}$ and adjoint matter fields $\Phi_a$, $r_{ab}$ denote the exact superconformal R-charge of $\Phi_{ab}$, with $r_{aa}$ denoting the R-charge of $\Phi_a$. In general, these R-charges are algebraic numbers~\cite{Intriligator:2003jj}. 

To proceed, we will now evaluate explicitly the functions $ \kappa(u_i^a-u_j^b  + \tfrac{r_{ab}}{2}) $ in the intervals $u_i^a-u_j^b \in (-1,1)$, focussing on a single chiral field, but bearing in mind the anomaly cancellation conditions, that hold only after considering globally  the complete field content of the theory. 
Below we will  denote the R-charge of the chiral field as $r$, and the variable $u_i^a-u_j^b$ as $x$. In order to study the function 
$\kappa (x+\frac{r}{2}$) in the interval $x \in (-1,1)$ it is convenient to divide this in sub-intervals as in (\ref{potforsym}), but presently 
$x\to -x$  is not a symmetry, and therefore the intervals will not be symmetric around $x=0$. 

For simplicity below we will make the additional assumption that the R charge of each chiral field is in the range $0<r<1$,  which holds for 
 infinitely many examples of  ${\cal N}=1$  superconformal quiver gauge theories. These include the Klebanov-Witten model \cite{Klebanov:1998hh}, orbifolds of ${\cal N}=4$ SYM, as well as the $Y^{p,q}$  infinite family \cite{Benvenuti:2004dy}. 
 We begin recalling  the definition of the integer part 
 \be
\lfloor x +\tfrac{r}{2}  \rfloor  = 
\begin{cases}
& -1   \qquad ~\; \text{for }~~~ x \in I_-\\
& \ 0  \qquad~~~\text{for } ~~~ x \in I_0  \\
& \ 1  \qquad~~~\text{for }   ~~~  x \in I_+  \\
\end{cases}
\ee
where we defined for convenience the three intervals $I_- = (-1, - \frac{r}{2})$, $I_0 = [- \frac{r}{2}  , 1- \frac{r}{2})$, and 
$I_+ = [1- \frac{r}{2}  , 1)$. It is useful to  define the function
 \be
\hat f(x) \equiv - \kappa ( x +\tfrac{r}{2} )  + \frac{r(r-1)(r-2)}{4} \, , 
\label{topolino}
\ee
which  is expanded out as  
 \be
- \hat f(x) = 
\begin{cases}
&  2 x^3+  \frac{3 r^2 x}{2}+\frac{3 r^2}{2}+3 r x^2+3 r x+3 x^2+x  \qquad  \qquad \qquad~~~ \text{for }  x\in  I_-   \\
 & 2 x^3+ \frac{3 r^2 x}{2}+3 r x^2 -3 x^2 -3 r x +x \qquad\qquad \qquad \qquad  ~~~~\; \text{for }   x\in I_0   \\
& 2 x^3+  \frac{3 r^2 x}{2}-\frac{3 r^2}{2}+3 r x^2 -9 x^2-9 r x+13 x +6 (r-1) \quad~\text{for }   x\in I_+
\label{fstarts}
\end{cases}
\ee
Summing  the constant term in (\ref{topolino}) over all the chiral fields in the quiver gives
 \begin{align}
\sum_{i,j} \sum_{ab}\frac{1}{4} r_{ab}(r_{ab}-1)(r_{ab}-2) &=
\frac{1}{4} \left( \mathrm{Tr}R^3 -  \mathrm{Tr}R\right)\, , 
 \end{align}
where we used the fact that the gaugini  do not contribute to this linear combination of 't~Hooft anomaly coefficients. 
Notice that our assumption  $0<r_{ab}<1$ implies that each term in the sum above is positive, so
 that $\frac{32}{9} (3c-2a)=\mathrm{Tr}R^3 -  \mathrm{Tr}R>0$,
 in agreement with \cite{Hofman:2008ar}.
The potential  reads
\begin{align}
V_2 &=    \frac{1}{4}(\mathrm{Tr}R -\mathrm{Tr}R^3)+      \sum_{ab} \sum_{i,j}^N\hat  f(u_{ij}^{ab} )_{r_{ab}} \, ,
\end{align}
where the subscript $r_{ab}$ indicates that the function depends on the R-charge $r_{ab}$.
Notice that the  functions  (\ref{fstarts}) are continuous  in $(-1,1)$, although they are not smooth at  the junctions $x=-r/2$ and $x=1-r/2$.

We will now take advantage of the gauge anomaly cancellation conditions, in order to simplify the form of the functions (\ref{fstarts}).
For example, the term $2x^3$ is unchanged across the interval $(-1,1)$ and is therefore the same for each chiral field contribution. We can then implement the cubic gauge anomaly cancellation condition 
\be
 \sum_{ab}\sum_{i,j}(u_i^a-u_j^b)^3=0 \, ,
\label{anoma3}
\ee
thus effectively removing  the terms $2x^3$ from (\ref{fstarts}).
Similarly, using the conditions 
\begin{align}
\sum_{ab}\sum_{i,j}(r_{ab}-1)^2 (u_i^a-u_j^b)=0 \, ,  \qquad\qquad  \sum_{ab}\sum_{i,j}(u_i^a-u_j^b)=0 \, ,
\label{anoma2}
\end{align}
we can subtract off  (\ref{fstarts})  terms proportional to $(r-1)^2x$ and $x$, respectively. It is important that these terms are subtracted  in \emph{all} the sub-intervals where $\hat f(x)$ is defined. We obtain the expression 
  \be
\tilde f(x)  \equiv \hat  f(x) \simeq - 3 (r -1)  x^2 +
\begin{cases}
&  - 6( x+\frac{r}{2})^2 \qquad \qquad    ~\, \text{for }  x\in  I_-  \\
&0 \qquad \quad \qquad  \qquad\qquad\text{for }  x\in  I_0    \\
& 6   ( x+\frac{r}{2} -1)^2 ~~ \qquad \quad\text{for } x\in  I_+ \\
\end{cases}
\label{simplestform}
\ee
where the symbol $\simeq $ here indicates  that the equality holds  after using the conditions (\ref{anoma3}) and (\ref{anoma2}), and the 
final form for the potential reads
\begin{align}
V_2 &=    \frac{1}{4}(\mathrm{Tr}R -\mathrm{Tr}R^3)+      \sum_{ab} \sum_{i,j}^N \tilde f(u_{ij}^{ab} )_{r_{ab}} \, .
\label{summa}
\end{align}
Furthermore, using the  ABJ anomaly cancellation condition $\sum_{ab}\sum_{i,j} (r_{ab}-1) (u_{ij}^{ab})^2 +  \sum_{a}\sum_{i,j} (u_{ij}^{aa})^2=0$, we can also write the potential in the form
\begin{align}
V_2  =\, \frac{{\rm Tr}R - {\rm Tr} R^3}{4} +6   \sum_{\alpha\in \Delta_+}(\alpha\cdot u)^2    + 3 \sum_{I\in {\rm chirals}}\sum_{\rho_I\in {\rm R}_I} \omega(\rho_I\cdot u + \tfrac{r_I}{2})\, ,
\end{align}
where 
\be
\omega (y) \equiv y (\left| y\right| -1)+ (y-1) (\left| y-1\right| -1)\, ,
\ee
thus making direct contact with (\ref{eyecontact}).

Each of the summands in (\ref{summa}) is defined in sub-intervals that depend on $r_{ab}$, so that the complete potential will be defined in an unwieldy set of intervals.  If the variables  $u_{ij}^{ab}$ were independent,   we could analyse  the potential 
   as we did in the simpler case of ${\cal N}=4$ SYM.   
  \begin{figure}
	\centering 
\includegraphics[width=8cm]{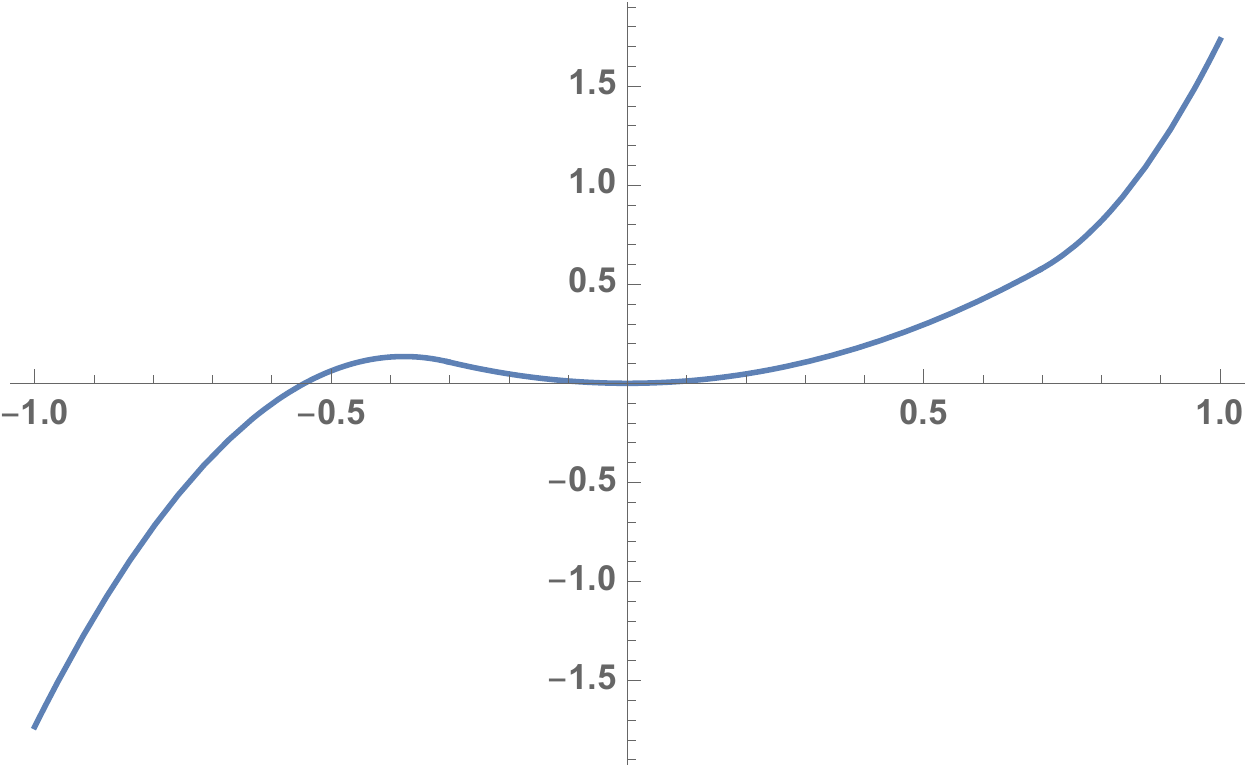}
\caption{The function $\tilde f(x)$ for a chiral field $Y$ in the  $Y^{2,1}$ quiver gauge theory \cite{Benvenuti:2004dy}, with 
 R-charge $r=\frac{1}{3} \left(3 \sqrt{13}-9\right)\simeq 0.605$.}
\label{skewshapef}
\end{figure}
    A generic example of function  $\tilde f(x)$ is plotted  in  Figure \ref{skewshapef}:
  since $\tilde f (x)$ has the dominant  saddle  point  at $x=0$,  this would then be the dominant saddle point of the full $V_2$. However, the  $u_{ij}^{ab} =u_i^a-u_j^b$ 
with $\sum_{i}u_{i}^{a} =0$, and a rigorous general analysis, without any further assumptions, appears to be complicated. 
 To proceed, below we will first restrict attention  to the classes of non-chiral theories, and then we will discuss more general classes of theories within 
  a simplifying ansatz for the gauge holonomies.  

We now specialize to the class of non-chiral theories, maintaining the assumption that all $R$-charges are less than 1.
Examples  include the Klebanov-Witten model, where  all the bi-findamentals have  R-charges equal to $r=1/2$; the ${\cal N}=2$,
$\mathbb{C}^2/\mathbb{Z}_2\times \mathbb{C}$ orbifold theory, where all the fields have R-charges $r=2/3$; the $L^{a,b,a}$ family \cite{Franco:2005sm,Butti:2005sw,Benvenuti:2005ja} 
 provides infinite examples of non-chiral quivers with bi-fundamentals and adjoints, where the R-charges of the various fields take different irrational 
values, all strictly less than 1.

Non-chiral quivers comprise only adjoint fields or pairs of bi-fundamentals in a  representation $\rho$ and its  charge conjugate  
$-\rho$, respectively.  Assuming that charge conjugated fields have equal R-charges,  we can write the potential (\ref{summa}) as
\begin{align}
V_2 &=   \frac{1}{4}(\mathrm{Tr}R -\mathrm{Tr}R^3)+     \sum^{}_{ab}{}' \sum_{i,j}^N \left(  \tilde f(u_{ij}^{ab} )_{r_{ab}} +  \tilde f(-u_{ij}^{ab} )_{r_{ab}}   \right)   \, ,
\end{align}
where now the primed sum is over all pairs of bi-fundamentals, or adjoints. For adjoint fields we define the primed sum with an extra factor of $\frac{1}{2}$, to avoid double counting. We can thus restrict attention to the function 
\be
 f(x) \equiv \tilde f(x)+\tilde f(-x) \, , 
\ee
and from (\ref{simplestform}) we find 
\be
 f(x)  = 
\begin{cases}
&- (r-1)6x^2 \qquad \qquad\qquad  \quad  \ \text{for } |x| \leq \frac{r}{2} \\
&\frac{3}{2} r \left(-4x^2 +4\left| x\right| -r\right) \qquad ~~~\,  \text{for }  \frac{r}{2}  < |x| \leq 1 - \frac{r}{2}   \\ 
& - (r-1) 6(|x|-1)^2  \qquad\qquad  ~\text{for }  1 - \frac{r}{2}  < |x| < 1 \\
\end{cases}
\label{noncpotentialab}
\ee
which is manifestly  a positive, even function of $x$. 
For a non-chiral theory with all R-charges less than 1 the complete potential thus reads 
\begin{align}
V_2 &=    \frac{1}{4}(\mathrm{Tr}R -\mathrm{Tr}R^3)  + \sum^{}_{ab}{}'   \sum_{i,j}^N  f(u_{ij}^{ab} )_{r_{ab}} \, ,
\label{dinnertime}
\end{align}
where the sum is over all pairs of bi-fundamentals stretching between nodes $a$ and $b$, or  fields in the adjoint representation, for which the contribution has an extra factor of $\frac{1}{2}$, as noted above.

From this we can  immediately recover the result  (\ref{potforsym}) for ${\cal N}=4$ SYM, taking $r_{a}=\frac{2}{3}$, $a=1,2,3$, and noting that the sum in (\ref{dinnertime}) is over all $i,j =1,\cdots ,N$.
For any $r_{ab}<1$ the shape of the potential is always a superposition of terms with the same shape as in Figure~\ref{functionf}; in particular, the absolute minimum is at $x=0$ and the two maxima at $x=\pm 1/2$. This proves that for ${\rm Re}\left(\frac{\ii}{\sigma\tau}\right)<0$, the Cardy-like limit of the index is dominated by the saddle where all gauge holonomies vanish.

Finally, let us come back to the more complicated case of chiral theories, namely quivers where not all bi-fundamental fields stretching from node $a$ to $b$ are accompanied by a field stretching from $b$ to $a$ and having the same R-charge. We notice that taking the
ansatz  that the gauge holonomies are equal for all the different nodes, i.e. 
\be
u^a_i = u_i \qquad  \forall a\, ,
\ee
and with all R-charges satisfying $r_{ab}<1$, then we can conclude that $u^a_i =0$ is the global minimum of  the potential within this ansatz. Indeed, with this ansatz we have $u^a_i-u^b_j = u_i-u_j\equiv u_{ij}$ and the potential \eqref{summa} can be written as\footnote{ Note that, in contrast to (\ref{dinnertime}), the sum over $ab$ here runs over all the bi-fundamentals and for adjoints the is no extra factor of $\frac{1}{2}$.}
\begin{align}
\label{newformu}
V_2 &=  \frac{1}{4}(\mathrm{Tr}R -\mathrm{Tr}R^3)  + \sum_{ab} \sum_{i<j}^N  f(u_{ij} )_{r_{ab}} \, ,
\end{align}
and for $r_{ab}<1$ the second term is a sum of non-negative contributions, with the only zero in $u_{ij}=0\ \Rightarrow\ u_i=0$. Although the ansatz $u^a_i = u_i$ is certainly not exhaustive for general quivers, it is well justified for particularly symmetric cases such as the orbifolds
of $\mathcal{N}=4$ SYM (which have $r_{ab}=2/3$),  corresponding to arbitrary  quotient singularities $\mathbb{C}^3/\Gamma$, with $\Gamma \in SU(3)$.
A particular class of ${\cal N}=2$ orbifolds was discussed in a related context in \cite{Honda:2019cio}.

\section{The large-$N$ saddle points of the superconformal index \label{Sec:largeN}}

In this section we analyze the saddle points of the integral representation~\eqref{index_other_writing} 
of the superconformal index in the large-$N$ approximation. We begin by writing the 
index in a representation suitable for analyzing the saddle points for real values of~$u$. 
As a check we verify explicitly that this representation is the same as the Hamiltonian version 
of the index. We present the saddle point equations in a large but finite-$N$ theory. 
We find a family of solutions to our saddle points equations for a class of~$\CN=1$ SCFTs.
This includes the solution $u=0$ which we found in the previous section in the Cardy-like limit,
which reproduces the black hole entropy. 

For ease of presentation we will write the index as follows, 
\be \label{IGammarel}
\mathcal I(\t,\s;n_0) 
\= \frac{1}{|\mathcal{W}|} \, \int  \prod_{i=1}^{\text{rank}(G)} \! d u_i  \; 
\prod_{I} \underset{\rho \in \text{R}_I}{\prod{}^{'}} \Gamma_e \bigl(z_I (u)| \t,\s \bigr) \,,
\ee
where we recall our definition
\be \label{zIdefagain}
z_I\=z_I(u) \=  \rho\cdot u+\frac{r_I}{2} \bigl(\t+\s-n_0\bigr) \,, \quad \text{ with } \rho\in \text{R}_I \,.
\ee
In this equation, and throughout this section, the index $I$ runs over all the multiplets of the theory.
This also includes the contribution of the vector multiplet which effectively contributes as a chiral multiplet 
of R-charge~$r=2$. In addition we treat the zero roots of the vector multiplet (corresponding to the
Cartan elements) in a special manner because they lead to zero modes as is well known. 
The prime on the product denotes that for the Cartan elements we remove the zero modes and 
thus effectively replace the elliptic gamma function by 
the prefactor containing Pochhammer symbols in 
Equation~\eqref{index_other_writing}, \eqref{ZvecZchi}. 

The representation of the elliptic gamma function that we will use in this section requires us to restrict to 
the range of parameters
\be \label{CondIntRep}
0<\Im(z)<\Im(\t)+\Im (\s) \,.
\ee
This condition combined with~\eqref{zIdefagain} implies that $0<r<2$. 
Therefore we cannot use it as it stands for the vector multiplet for which~$r=2$. To avoid this 
we deform the theory slightly by taking the R-charge of the vector multiplet to be~$2-\ve$, and correspondingly
deform all the other R-charges of the matter multiplets so that the anomaly cancellation condition still holds. 
We then define the values of the integrals to be the limiting value as~$\ve \to 0$. 
This regulator is related to the subtlety about the point~$u=0$ that was discussed near Equation~\eqref{estimate_vectors} 
in the previous section. We will clarify this further  below.

In the range of parameters~\eqref{CondIntRep}  
the elliptic gamma function can be rewritten in the following manner~\cite{Felder}, \cite{Narukawa},
\bea \label{rep1}
\Ge \left(z;\t,\s\right)  &\=&  \exp \bigl(-S_1(z;\t,\s) \bigr) \,, \\
S_1(z;\t,\s) & \= &  \frac{\i}{2}  \sum_{n=1}^\infty \, 
	\frac{\sin \bigl(\pi n (2 z - \t - \s) \bigr)}{n  \,\sin  (\pi \t n ) \, \sin (\pi \s n)}  \,.
\label{defS1}
\eea
Using this representation we can write the index~\eqref{IGammarel} as 
\be  \label{ISrel}
\mathcal I(\t,\s;n_0) 
\= \frac{1}{|\mathcal{W}|} \, \int \Bigl( \prod_{i=1}^{\text{rank}(G)} d u_i  \, \Bigr)\; \exp \bigl(-S(u) \bigr)  \,,
\ee
where the \emph{effective action} of~$u\equiv\{u_i \}$ is
\be \label{actionS}
S(u) \=  {\underset{I, \, \rho \, \in \, R_I}{\sum{}^{'}}}  \, S_1(z_I(u);\t,\s) \,,
\ee
where the prime on the sum means that we treat the contribution of the Cartan elements in the vector multiplet in a special manner as above.
In particular this means that their contribution is independent of~$u$.
Using the representation~\eqref{defS1} we can write the effective action as
\be \label{SPEform}
S(u) \= \frac{\i}{2} \sum_{n=1}^\infty \, 
\frac{1}{n} \, \frac{1}{\sin  (\pi \t n ) \, \sin (\pi \s n)} \,  {\underset{I, \, \rho \, \in \, R_I}{\sum{}^{'}}}  \, \sin\bigl( 2 \pi n\,( \rho\cdot u + \d_I) \, \bigr)  \,,
\ee
with
\be \label{defdelta}
\d_I \= \frac{1}{2} (r_I-1)(\t+\s) - \frac{1}{2}r_I \, n_0 \,,
\ee
or, equivalently, 
\be \label{SPEform1}
S(u) \= \frac{\i}{2} \sum_{n=1}^\infty \,
\frac{1}{n} \, \frac{1}{\sin  (\pi \t n ) \, \sin (\pi \s n)} \; \frac{ g(nu, n\t,n\s,nn_0-\tilde{g}(nu, n\t,n\s,nn_0) }{2\ii} \,,
\ee
with
\be \label{defg}
g(u,\t,\s,n_0) =  {\underset{I, \, \rho \, \in \, R_I}{\sum{}^{'}}}\!\! \exp\bigl( 2 \pi \i \,( \rho\cdot u + \d_I) \, \bigr)\,,\qquad \tilde{g}(u,\t,\s,n_0) =  {\underset{I, \, \rho \, \in \, R_I}{\sum{}^{'}}}\!\! \exp\bigl( -2 \pi \i \,( \rho\cdot u + \d_I) \, \bigr) \,.
\ee

The above formula for the index is essentially the same one studied in~\cite{Kinney:2005ej, Romelsberger:2005eg} 
for~$\CN=1$ SCFTs that can be reached via a Hamiltonian trace formula. 
Our function~$g(u)$ is related to the ``single-letter trace", and the right-hand side 
of~\eqref{SPEform1} is the plethystic exponential acting on this trace. 
In order to demonstrate this explicitly, let's consider the case of~$U(N)$ gauge group and adjoint matter. 
In this case we have 
\be\begin{split} \label{defgun}
g(u) & \=  
\sum_{I} \,  \rme^{2\pi \i \d_I} \underset{j,k=1}{\sum^{N}{}^{'}}  \, \rme^{2 \pi \i  (u_j-u_k)}  
\= \sum_{I} \,   \rme^{2\pi \i \d_I} \; \underset{j,k=1}{\sum^{N}{}^{'}} \, \cos \bigl(2 \pi  (u_j-u_k) \bigr) \,, \\
\Rightarrow \, \frac{g(u) - \tilde{g}(u)}{2\ii} & \= \sum_{I} \,   \sin(2\pi  \d_I) \; \underset{j,k=1}{\sum^{N}{}^{'}}\, \cos \bigl(2 \pi  (u_j-u_k) \bigr) \,.
\end{split}
\ee

The field content of $\mathcal{N}=4$ SYM in our~$\CN=1$ language consists of a vector multiplet with\footnote{We suppress our regulator~$\ve$ 
here for the purposes of comparison.} 
R-charge $r=2$ and 
three adjoint chiral multiplets with R-charge $r=\frac{2}{3}$. The sum over~$I$ in~\eqref{defgun} clearly factors out of the sum over~$i,j$,
and after a short calculation we reach 
\be \label{sumIdelta}
\sum_{I} \,  \frac{\sin(2\pi n \d_I)}{\sin{(\pi n \t)}\sin{(\pi n \s)}}  \= -2 \i \,\frac{ (1-w^n)^3}{(1-a_1^n) (1-a_2^n)} \,,
\ee
with
\be
a_{1} \= \exp{ \bigl( 2\pi \i  \t \bigr)} \,,   \quad a_{2} \= \exp{ \bigl( 2\pi \i  \s \bigr)} \,, 
\quad w \=  \exp{\Bigl( \, \frac{2 \pi i}{3}  \bigl(\t+\s-n_0 \bigr) \, \Bigr)} \,.
\ee
The effective action \eqref{SPEform1} can be now written as 
\bea\label{SPEformKMMR}
S(u) \=\underset{j,k=1}{\sum^{N}{}^{'}}\, \sum_{n=1}^\infty \,\frac{1}{n} \,
\frac{ (1-w^n)^3}{(1-a_1^n) (1-a_2^n)} \, \cos \bigl(2 \pi n  (u_j-u_k) \bigr) \,.
\eea
This is the generalization to arbitrary~$n_0$ of the effective action studied in~\cite{Kinney:2005ej} in the case when the 
chemical potentials of the three R-charges are equal.

\subsection{The continuum matrix model and its saddle points at large~$N$}

The above representation of the effective action contains~$O(N^2)$ terms in the sum over the weights~$\rho$,  
and so we can use the saddle point approximation to evaluate the integral~\eqref{IGammarel}  at large~$N$. 
To see this more precisely, we write the index as an integral over~$N \times N$ matrices~$U$~\cite{Aharony:2003sx},  
\be
\mathcal I(\t,\s;n_0) 
\= \frac{1}{|\mathcal{W}|} \, \int \, [dU]\, \exp \biggl( \, \sum_{n=1}^\infty \, 
\frac{1}{n} \, \sum_{I} \, z^{R_I}(n\t,n\s,nn_0) \, \chi_{R_I} (U^n) \, \biggr) \,,
\ee
where~$\chi_{R}$ is the character defined as the trace of the group element~$U$ in the representation~$R$,
and~$z^R$ is the single-letter index trace in the representation~$R$. For~$U(N)$ gauge groups with adjoint matter
$\chi_{R_I} (U^n) = \text{tr} \, U^n \, \text{tr} \, (U^\dagger)^n$. Similarly if we have product gauge groups and 
bi-fundamental matter we will have a product of two traces in the action. These are the types of theories we analyze in this 
paper. The product of two traces leads to an overall~$N^2$ in front of the action, so that at leading order in the large-$N$
approximation we simply have to extremize the action. 

One can analyze the large-$N$ limit, as in~\cite{Aharony:2003sx,Kinney:2005ej,Dolan:2008qi}, by promoting~$u_i$
to a continuous variable~$u(x)$ and replacing the sum over~$i$ by an integral over~$x$. 
We can further replace the integral over~$x$ by an integral over~$u$ with a factor of the density~$\mu(u) = \frac{dx}{du}$
which obeys a normalization condition.\footnote{In order to avoid any confusion, we note that in our notation~$\rme^{2 \pi \i u}$ labels  
eigenvalues of the matrix~$U$. This leads to a Jacobian factor which has already been taken into account while 
writing the integral~\eqref{IGammarel}~\cite{Cabo-Bizet:2018ehj}. Also the density of eigenvalues is often denoted by~$\rho(u)$, but 
we have already used~$\rho$ to denote the weights of the representations so we use~$\mu(u)$ for the eigenvalue density.}
In this limit, we obtain the following effective action, 
\be
S \= N^2 \int \, du \, dv \, \mu(u) \, \mu(v) \, V(u-v) \,,
\ee
with the pairwise potential~$V$ taking the form 
\be
V(u) \= \sum_{n=1}^\infty \,\frac{1}{n} \, V_n (u) \,,
\ee
where the value of~$V_n$ can be read off, for example, from the action~\eqref{SPEform}. For the case of~$\CN=4$ SYM 
one has
\be
V_n \= \frac{ (1-w^n)^3}{(1-a_1^n) (1-a_2^n)} \, \cos \bigl(2 \pi n  u \bigr) \,.
\ee
The saddle point equations for~$u_i$ can be written in the continuum representation as follows,
\be \label{SPEqncont}
\int \, dv \, V'(u-v) \, \mu(v) \= 0\,.
\ee
This equation was interpreted in~\cite{Aharony:2003sx} as a balancing condition for an extra eigenvalue in the equilibrium configuration.

Now we move to the analysis of the saddle point equations. 
Our effective action is completely governed by the function~$g(u)$ as shown in~\eqref{SPEform1}. 
From this representation we see that the saddle point equations~\eqref{SPEqncont} are solved by
\be \label{gEqncont}
g'(u) \; \equiv \; \frac{\delta g}{\delta u} \= \int \, dv \, V_1'(u-v) \, \mu(v) \= 0\,.
\ee
Let us focus for now on the adjoint representation for which~$g$ is given in~\eqref{defgun} 
(after promoting the sum over~$j,k$ to integrals).  
Using the fact that the sine function has vanishing zero mode, it is clear in this case that the uniform 
distribution~$\mu(u)=\mu_0$ is a solution of the equation~\eqref{gEqncont}.
At the other extreme, the distribution~$\mu(u) = \delta(u)$ is also a solution to the saddle point equations
because the sine function vanishes at the origin. An analysis of the competition between these types of 
saddles was performed in~\cite{Aharony:2003sx} in the context of the partition function of SYM at finite temperature. 
For the case of the index, when the chemical potentials are restricted to be purely imaginary, 
the paper~\cite{Kinney:2005ej} showed that the absolute minimum of the effective potential is zero, 
and it is achieved by the uniform distribution, 
thus bypassing the need to analyze all saddle points. For the case of arbitrary chemical potentials on the other hand, 
we have to look for other saddles, and that is what we now turn to.

In fact we go back to the original discrete problem and analyze the saddles for real values of~$u$. As we shall 
see there are many saddles which interpolate between the two mentioned above. We note, however, 
that our saddle point solutions can be translated back into the continuum language as done above for 
the two extreme saddles.

\subsection{Large-$N$ saddle points of the discrete model \label{Sec:SaddlePoints}}

In our original variables~$u_i$, the saddle point equations for a generic theory are given by
\be
\p_{u_j} S(u) \= 0 \,,\qquad j=1,\,\dots , N\,.
\ee
Recalling that~$u$ is real variable, we see that these equations are solved by 
\be \label{defgj}
g_j(u) := \p_{u_j} g(u) \= 0 \,, \qquad j=1,\,\dots , N\,.
\ee
(The specific form of~$g$ in~\eqref{defg} ensures that a solution to~$g_j(u,\t,\s,n_0)=0$
is also a solution to $g_j(nu,n\t,n\s,nn_0)=0$ for~$n \in \IZ$.)
In this paper we only focus on solutions to~\eqref{defgj} and do not attempt to classify all the 
saddle points of the effective action.
As we shall see in the following, we find a particular solution to~\eqref{defgj} for which the index~\eqref{IGammarel} 
has an asymptotic growth of states that equals the entropy of the black hole in the Cardy-like limit.

\subsubsection*{$U(N)$ gauge group}

Let us begin with the case that the gauge group is~$U(N)$ and all matter is in the adjoint representation 
(this of course includes the case of~$\CN=4$ SYM). In this case~$g(u)$ is given by~\eqref{defgun}
and its derivatives are given by\footnote{The prime on the summation symbol can be dropped because 
the Cartan elements give a contribution independent of~$u$ and therefore do not contribute to the derivative.}
\be\begin{split} \label{defpgun}
g_j(u) & \=\ 2 \pi \i \sum_{I} \,  \rme^{2\pi \i \d_I} \sum_{k=1}^N \, \bigl( \rme^{2 \pi \i  (u_j-u_k)} - \rme^{2 \pi \i  (u_k-u_j)} \bigr) \\
& \= - 4 \pi \sum_{I} \,  \rme^{2\pi \i \d_I} \sum_{k=1}^N \, \sin \bigl(2 \pi  (u_j-u_k) \bigr)\,.
\end{split}
\ee
The saddle point equations are thus solved by the solutions to 
\be \label{SadAux}
\sum_{k=1}^N \, \sin\bigl( 2 \pi (u_{j} - u_{k}) \bigr)  \= 0 \,, \qquad j=1,\dots,N\,.
\ee

We now look for solutions to the equations~\eqref{SadAux}. Clearly the configuration~$u_j=0$, $j=1,\dots, N$ is a solution.
This corresponds to a distribution in which all the eigenvalues are clumped together at one point. 
Using the fact that the $N^\text{th}$ roots of unity add up to zero, it is also easy to see that the configuration~$u_j= \frac{j}{N}-\frac12$, 
$j=1,\dots,N$, solves the system~\eqref{SadAux}. This latter solution corresponds to a uniform distribution of eigenvalues in the unit interval.
There are also other solutions which lie between these two extremes. 
Thinking of~$2 \pi u_i$ as angles, we can plot them on a circle of unit radius.
A configuration in which the variables~$u_{i}$ coincide with the vertices of a regular polygon inscribed in this circle
with an equal number of them at each vertex of the polygon is a solution to the equations~\eqref{SadAux}.  

\begin{figure} 
\begin{center}
\includegraphics[width=0.3\textwidth]{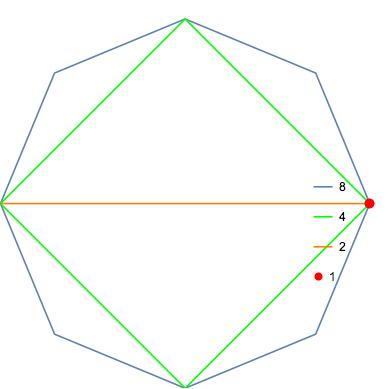}
\end{center}
\caption{$N=8$ $K$-gon solutions with $K=8,4,2,1$. 
The vertices represent clusters of $N/K$ eigenvalues in the complex plane~$\rme^{2\pi \i u}$.}
\label{kgon}
\end{figure}

More precisely, choose a divisor~$K$ of~$N$, and consider the set of points 
\be \label{kasols}
P_K \= \Bigl\{ \,\frac{j}{K} + c_K \,, \; \; j= 1,\dots, K  \, \Bigr\}\,,
\ee 
where~$c_K$ is a real constant (which can be chosen so as to bring all the values of~$u$ to the range~$(-\frac12, \frac12]$).
In the complex plane~$w=\rme^{2 \pi \i u}$ these~$K$ points label the vertices of a regular $K$-gon (see Figure~\ref{kgon} for an example
with~$N=8$ and $c_K=0$).
It is easy to check that the configuration where all the variables~$u_{i}$ take values in~$P_K$
with an equal number~$\frac{N}{K}$ of them at each vertex of the polygon solves the equations~\eqref{SadAux}. 
The solutions with larger~$K$ describe a more spread out distribution of eigenvalues. In particular, 
the solution with~$K=N$ corresponds to a uniform distribution of eigenvalues and~$K=1$
correspond to a configuration when all the eigenvalues clump together at one point, as discussed above.

We discuss the action of the various solutions in more detail in the next subsection. 
%The result is that the action of the completely clumped solution~$u=0$ is proportional to~$N^2$ and is the dominant
%solution, 
%while the action of the partially clumped or uniform distribution is proportional to~$N^2/K$ and is therefore subleading,  
%in the large-$N$ limit.
The result is that the completely clumped solution~$u=0$ is the dominant
solution, while the partially clumped or uniform distribution is subleading 
in the large-$N$ limit.

\subsubsection*{Quiver theories}

Now we move to theories with gauge groups~$G = \prod_{a=1}^\nu U(N_a)$ and find the saddle points as above.
We take each~$N_a$ to be large. We further assume that 
the matter is in the adjoint or bi-fundamental representations of the gauge group such that the theory is anomaly-free.
Within these assumptions we will find a set of saddle points, similar to the above analysis for~$U(N)$ gauge groups. 
The quiver diagram has~$\nu$ nodes, and at each node~$a$ we have the gauge group~$U(N_a)$.
We label the Cartan elements of the gauge fields as~$u_j^a$, $a=1,\dots, \nu$, $j=1,\dots,N_a$. 
The matter multiplets are described by arrows connecting pairs of modes~$(a,b)$. 
The nodes~$a$ and~$b$ are connected by~$\t_{ab}$ arrows
pointing from~$a$ to~$b$ and~$\t_{ba}$ arrows from~$b$ to~$a$. 
Each arrow represents one chiral multiplet in a bi-fundamental representation of~$U(N_a) \times U(N_b)$
with R-charge~$r_{ab}$.

The function~$g$ is now given by 
\be \label{gquiver}
g(u) \=  \sum_{a,b=1}^\nu \,  \t_{ab} \, 
\rme^{2\pi \i \d_{ab}} \,  \underset{j=1}{\sum^{N_a}{}^{'}} \, \underset{k=1}{\sum^{N_b}{}^{'}} \, \rme^{2 \pi \i  (u^a_j-u^b_k)}  \,,
\ee
with
\be
\d_{ab} \= \frac{1}{2} (r_{ab}-1)(\t+\s) - \frac{1}{2} \, r_{ab} \, n_0 \,,
\ee
and its derivatives~$g_{ja}(u)  := \p_{u_j^a} \, g(u)$ are given by
\be\begin{split} \label{gjquiver}
\frac{1}{2\pi \i} \, g_{ja}(u) 
&   \= \t_{aa} \, \rme^{2\pi \i \d_{aa}} \, \sum_{k=1}^{N_a} \, \sin({2 \pi  (u^a_j-u^a_k)}) \; + \\
&  \qquad \qquad + \; \sum_{b\neq a} \,  \biggl( \, \t_{ab} \, \rme^{2\pi \i \d_{ab}} \, \sum_{k=1}^{N_b} \,   \rme^{2 \pi \i (u^a_j-u^b_k) } 
\; -  \; \t_{ba} \, \rme^{2\pi \i \d_{ba}} \,  \sum_{k=1}^{N_b} \,   \rme^{2 \pi \i  (u^b_k-u^a_j)}  \, \biggr) \,.
\end{split}
\ee

We show in Appendix~\ref{App:Saddle} that these equations admit solutions with partial or no clumping behaviour 
for a generic quiver. The action of these solutions vanishes at large~$N$ exactly as in the~$U(N)$ case. 
In order to check whether the completely clumped configuration~$u=0$ is indeed a solution,
we now specialize to a couple of cases. 

\paragraph{Non-chiral quivers.} 
This is a class of theories in which each arrow from~$a$ to~$b$ is accompanied by an arrow from~$b$ to~$a$. 
In this case~$\t_{ab} = \t_{ba}$ and~$\d_{ab} = \d_{ba}$, which leads to
\be \label{gnonchir}
\begin{split}
g(u) & \=  \sum_{a,b=1}^\nu \,  \t_{ab} \, 
\rme^{2\pi \i \d_{ab}} \, \underset{j=1}{\sum^{N_a}{}^{'}}
\underset{k=1}{\sum^{N_a}{}^{'}}  \, \cos \bigl({2 \pi  (u^a_j-u^b_k)}\bigr) \,,\\
g_{ja}(u)   & \= - 4 \pi \sum_{b=1}^\nu \,  \t_{ab} \, 
\rme^{2\pi \i \d_{ab}} \,  \sum_{k=1}^{N_b} \, \sin \bigl({2 \pi  (u^a_j-u^b_k)}\bigr) \,.
\end{split}
\ee 
Our solutions from the~$U(N)$ case carry over easily.
In particular, the completely clumped configuration~$u^a_j=0$, $j=1,\dots, N_a$, $a=1,\dots, \nu$, 
clearly solves the saddle points.

\paragraph{Chiral quivers with equal R-charges.} 
In the case when the R-charges of all the matter fields are equal (the orbifold theories~$Y^{p,0}$
and~$Y^{p,p}$ fall into this category), it follows from~\eqref{defdelta} 
that~$\d_{ab}=\d$, a constant. In that case, the configuration~$u_{ia}=0$
is a solution to the equations~$g_{ia} = 0$. To see this, note that in this case  
\be
g_{ia}(u = 0) \=  \rme^{2\pi \i \d} \, 
\sum_{b\neq a} \,  \bigl( \, \t_{ab} \; -  \; \t_{ba}   \, \bigr) N_b \,,
\ee
which vanishes due to anomaly cancellation in the quiver theory.

\paragraph{Other gauge groups.} 
In all the above analysis we have discussed~$U(N)$ gauge groups, but a similar analysis can be done 
for other gauge groups with appropriate modifications. In particular, the configuration~$u=0$ is a saddle point
for all gauge groups and, for a theory where~$U(N)$ is replaced by~$G$, the action at the saddle point is of 
the same form as~\eqref{Su0Un},  with the replacement~$N \to \text{rank}(G)$.

For example, the effective actions of $\mathcal{N}=4$ SYM 
with gauge groups $SU(N)$, $SO(N)$, $Sp(2 N)$ are as follows. 
Using the usual notation of the classification of the corresponding algebras 
$A_N=su(N+1)$, $B_N=so(2N+1),\, C_N=sp(2 N)$ and $D_N=so(2N)$ ($N>3$), we have
\bea\label{efAffaction}
S_{G}(u)\=\frac{\i}{2} \, \underset{I}\sum{}^{'} \, \sum_{n=1}^{\infty} 
\frac{\sin \bigl( 2 \pi \delta_I n \bigr) }{n \sin{ (\pi  \t n )}\sin{ (\pi  \s  n )}} \, \Xi_G (n u), 
\eea
with
\bea
\Xi_{A_N}(nu)&=& 2 \sum_{i<j=1}^{N+1} \cos{\bigl( 2 \pi n (u_i-u_j)\bigr)},\qquad \sum_i^{N+1} u_i=0,
\\
\Xi_{B_N}(nu)&=& 2 \sum_{i<j=1}^N \!\left( \cos{\bigl(2 \pi n (u_i+u_j)\bigr)} +  \cos{\bigl(2 \pi n (u_i-u_j)\bigr)}  \right) + 2\sum_{i=1}^N \cos{\bigl( 2 \pi n  u_i \bigr)},\ \qquad \label{eq:XiBN}
\\
\Xi_{C_N}(nu)&=& 2 \sum_{i<j=1}^N \!\left( \cos{\bigl(2 \pi n (u_i+u_j)\bigr)} +  \cos{\bigl(2 \pi n (u_i-u_j)\bigr)}  \right) + 2\sum_{i=1}^N \cos{\bigl( 4 \pi n u_i  \bigr)},\ \qquad \\
\Xi_{D_N}(nu)&=& 2 \sum_{i<j=1}^N \!\left( \cos{\bigl(2 \pi n (u_i+u_j)\bigr)} +  \cos{\bigl(2 \pi n (u_i-u_j)\bigr)}  \right) .\label{eq:XiDN}
\eea
We checked that %our~$K$-gon solutions, $K \ge 1$, are real saddles of all such effective actions $S(u)$. In particular 
$u=0$ %($K=1$)
 is a saddle of all such effective actions $S(u)$.\footnote{We thank Antonio Amariti for pointing out a missing term in Eqs.~\eqref{eq:XiBN}--\eqref{eq:XiDN} as well as a wrong claim on $K$-gon solutions for non-unitary groups, that appeared in a previous version of this paper.}
For the~$SU(N)$ theory all our~$K$-gon solutions, $K \ge 1$, are saddles. To see this, one starts from the integrand of a $U(N)$ theory plus a term proportional to the Lagrange multiplier $\Lambda$ 
that multiplies the ``trace" $\sum_{i=1}^{N} u_i$. The variation of the effective action with respect to $\Lambda$ imposes the tracelessness 
constraint. The saddle-point conditions for $u_i$ with $i=1,\ldots, N$ are satisfied by our $K$-gon ansatze with $K|N$ and the choices
\be
\Lambda\=0 \,, \text{   and   }  \, c_K\=-\frac{K+1}{2K} \,.
\ee

\subsection{Saddle-point evaluation of the integral and the Cardy-like limit \label{Sec:CardyLargeN}}

Now we discuss the effective action evaluated at the saddle points that we found in the previous subsection. 
For the~$U(N)$ theory,  
we see from~\eqref{defgun} that the value of the function~$g(n u)$
at the saddle points labelled by~$K$ is proportional to
\be
\underset{j,k=1}{\sum^{K}{}^{'}}  \, \rme^{2 \pi \i \,n  (j-k)/K} \,.
\ee
Since the contribution of the Cartan elements is suppressed by a factor of~$1/N$ compared to the 
full summation, we can use the full summation at leading order in large~$N$. The full summation can be 
evaluated easily to be
\be
\sum_{j,k=1}^K \, \rme^{2 \pi \i \, n (j-k)/K} \=  K^2 \sum_{c\in \IZ}\delta_{n,cK} \,.
\ee
This allows us to write the effective action, at leading order in large~$N$, of the solutions with~$K>1$
in terms of the solution with~$K=1$.
For the case~$K=1$ we simply have $u=0$, which implies that, in the large-$N$ limit,
\be
g(0,\t,\s,n_0) \=  {\underset{I, \, \rho \, \in \, R_I}{\sum{}^{'}}} \, \mathrm{e}^{2 \pi \i \, \d_I}
\= N^2 \sum_{I} \, \mathrm{e}^{2 \pi \i \, \d_I } \,.
\ee
Using~\eqref{SPEform}, \eqref{defdelta}, this leads to the effective action, in the large-$N$ limit, 
\be \label{Su0Un}
S \= \frac{\i}{2} \, N^2 \, \sum_{n=1}^\infty \,
\frac{1}{n} \, \sum_{I} \frac{\sin\bigl(\pi n ((r_I-1)(\t+\s) - r_I \, n_0 ) \bigr) }{\sin  (\pi \t n ) \, \sin (\pi \s n)}  \,,
\ee 
while for generic~$K$ we have 
\be\label{Su0UnK}
S^{(K)} \,=\, \frac{\i}{2} \, \frac{N^2}{K} \, \sum_{n=1}^\infty \,
\frac{1}{n} \, \sum_{I} \frac{\sin\bigl(\pi n \, K\, ((r_I-1)(\t+\s) - r_I \, n_0 ) \bigr) }{\sin  (\pi \t n \, K) \, \sin (\pi \s n\, K)}  \,.
\ee
Thus we see that the~$K>1$ solutions are suppressed by~$\rme^{(1-\frac{1}{K})N^2}$ compared to the leading~$K=1$
solutions.
For the quiver theories we can do a similar analysis to show that the action has a form which is quadratic in the~$N_a$.
In the case where the ranks are all equal, the action takes a form similar to~\eqref{Su0Un}
with an overall factor of~$N^2$.

Our result so far is that the effective action at the saddle point~$u=0$ is of the form~\eqref{Su0Un}.
It is important to note that we have only analyzed the saddles at real values of~$u$ so far. In fact we find that there are also 
other saddles with complex values of~$u$. 
For generic values of chemical potentials~$\t,\s$, we find that the black hole entropy is reproduced
by a saddle with non-zero imaginary part. These conclusions seem to go along the lines of the results  
of~\cite{Benini:2018ywd} obtained by analyzing a different set of equations coming from the Bethe-ansatz approach. 
We will present the details of this analysis in a forthcoming publication.

Now we proceed by going to the Cardy-like limit~$\t,\s \to 0$ on top of the large-$N$ limit.
%In that limit we see clearly that the solution with~$u=0$ is the leading saddle point in the large-$N$ limit among those found, 
%We can use the formulas of the previous section to recover the black hole entropy. 
%In this limit we show that we reach the formula~\eqref{result_integrand_divergent}, but now we only keep the 
%leading large-$N$ contribution.
In the method used in Section~\ref{sec:Cardy}, recall that we had to treat the point~$u=0$
carefully because it is a zero of the integrand of~\eqref{index_other_writing} (because of the contribution of the 
vector multiplet which has~$r=2$). 
The way we dealt with this in Section~\ref{sec:Cardy} is to estimate the integrand with confidence outside 
a small hole of size~$\upsilon$ around~$u=0$ (as prescribed by the theorem of~\cite{Rains:2006dfy} that we use).
Using this estimate we find a function which is highly peaked near~$u=0$ for distances larger than~$\upsilon$.
Although the actual function vanishes at~$\upsilon=0$, the uniformly converging estimate of~\cite{Rains:2006dfy} 
tells us that the integral is actually dominated by the limiting value of the function as we approach~$u=0$.
In this section we have used a different regulator, by explicitly deforming the R-charges away from 2.
Although our concern was the applicability of the representation~\eqref{rep1} we note that these two issues are 
related---we can see this explicitly by noting that the sum~\eqref{defS1} diverges at~$r=2$ and~$u=0$.
This gives us confidence in our regulator in order to evaluate the Cardy-like limit.

As we now show we can also reach the leading singular behavior in the Cardy-like limit,
to which we will refer as the \emph{extreme Cardy-like limit}, directly from the action~\eqref{Su0Un}
for the leading saddle point.
In order to see this we note that, up to the overall factor of~$N^2$, 
the action~\eqref{Su0Un} is a sum over the different multiplets of the infinite 
sum~$S_1$ written in~\eqref{defS1}, with the~$z$ variable fixed by the R-charge of the multiplet. 
In this limit the sum reduces to 
\bea 
S_1(z;\t,\s) & \= &  \frac{\i}{2}  \sum_{n=1}^\infty \, 
	\frac{\sin \bigl(\pi n (2 z - \t - \s) \bigr)}{n  \,\sin  (\pi \t n ) \, \sin (\pi \s n)} \,, \nn \\
&\; \longrightarrow \; & \frac{\i}{2} \sum_{n=1}^\infty \,  \frac{\sin (2 \pi  n  z )}{\pi^2 \, n^3 \, \t \s } 
\, \quad \text{as~$\t,\s \to 0$.}
\label{S1series}
\eea
Note that the right-hand side of~\eqref{S1series} converges if and only if~$z$ is real.
In the extreme Cardy-like limit we have precisely this situation. Indeed the relevant values of~$z_I$ 
in the limit reduce to   
\be
z_I(\rho, u=0)  \=  \frac{r_I}{2} \bigl(\t + \s-n_0\bigr) 
\; \longrightarrow \;  - \frac{r_I n_0}{2}   \,.
\ee

Now we use the formula
\be \label{BerFour}
\sum_{n=1}^\infty \,  \frac{\sin (2 \pi  n z)}{n^3 } \= \frac{2 \pi^3}{3} B_3(\{z \}) \,, \qquad z \in \IR \,,
\ee
where~$\{ z \} = z - \lfloor z \rfloor$ is the fractional part of~$z$ as before, and~$B_{3}$ is the Bernoulli polynomial 
\be
B_{3}(z) \=\bigl(z-\frac{1}{2} \bigr)^{3}-\frac{1}{4} \bigl(z-\frac{1}{2} \bigr) \=  z^3 - \frac{3 z^2}{2} + \frac{z}{2} \,.
\ee
The formula~\eqref{BerFour} can be proved in an elementary manner by calculating the 
Fourier expansion of the polynomial~$B_3$ for the range~$z\in [0,1]$ and then extending it 
by the periodicity of the left-hand side.\footnote{The Bernoulli polynomial also obeys the identity
\be 
-\frac{(2\pi \i)^3}{3!}B_3(z)  \;= \;  \text{Li}_{3}(\rme^{2\pi \i z})-\text{Li}_{3}(\rme^{-2\pi \i z}) \,.
\ee
Upon writing the sine function in the series~\eqref{S1series} as a difference of two exponentials, 
one gets two series which formally agree with those for~$\text{Li}_{3}(\rme^{2\pi \i z})$ and~$\text{Li}_{3}(\rme^{-2\pi \i z})$,
and it is tempting to try to reach~\eqref{BerFour} in this manner. 
However, this series representation for the polylogarithm only holds for~$z$ in the upper half-plane and therefore at most 
one of the two series converges. It is not clear to us how to proceed along that route. The steps we take 
to go from~\eqref{S1series} to \eqref{Bernoulli} avoid this problem as long as we stay in the extreme Cardy-like limit.} 
Thus we obtain, in the extreme Cardy-like limit,  
\be\label{Bernoulli}
\log \Ge \bigl(z_I,\t,\s  \bigr) \; \longrightarrow \; 
-\frac{\pi \i}{3 \, \t\s}  B_{3} \Bigl( \bigl\{-\frac{r_I n_0}{2} \bigr\} \Bigr)   \,,
\ee
which is consistent with~\eqref{estimate_Rains} (note that~$\kappa(x)=2B_3(\{x\})$), thus giving the total action
\be%\label{Bernoulli}
S \; \longrightarrow \;  -N^2 \frac{\pi \i}{3 \, \t\s}  \sum_I B_{3} \Bigl( \bigl\{-\frac{r_I n_0}{2} \bigr\} \Bigr)   \,.
\ee
Noting that the order of the Weyl group obeys~$\log |\mathcal{W}| = N \log N$ which is subleading in~$N$ compared 
to the leading~$N^2$ contribution, this leads to the black hole entropy at large~$N$. 
Similarly, using the action~\eqref{Su0UnK}, we find that the action for the $K$-gon saddles in this limit is
\be%\label{Bernoulli}
S^{(K)} \; \longrightarrow \;  -\frac{N^2}{K^3} \frac{\pi \i}{3 \, \t\s}  \sum_I B_{3} \Bigl( \bigl\{-\frac{K r_I n_0}{2} \bigr\} \Bigr)   \,.
\ee
Note that the factor of~$K$ inside the argument of the periodic function~$B_3(\{x\})$ does not contribute to the scaling of the growth
with~$K$.

\section*{Acknowledgements}
We would like to thank Antonio Amariti, Francesco Benini, Nadav Drukker, Alfredo Gonzalez, James Lucietti, and Leopoldo Pando Zayas for useful discussions.
The research of A.~C.~B.~and S.~M.~is supported by the 
ERC Consolidator Grant N.~681908, ``Quantum black holes:~A microscopic window into the microstructure of gravity''. 
The research of S.~M.~is also supported by the STFC grant ST/P000258/1.

\appendix

\section{Saddle points with partial or no clumping for generic quivers \label{App:Saddle}}

In this appendix we show that the generic quiver theories discussed in Section~\ref{Sec:SaddlePoints} have 
saddles with partial or no clumping behaviour, as in the~$U(N)$ case. 

It is convenient to first set up some notation and summarise a useful property of the exponential function.  
We define the set 
\be\label{PK}
P_K \=\Bigl\{ \, \frac{j}{K}+c_K \,, \; \; j \=1,\ldots, K \, \Bigr\} \,,
\ee
with $c$ an arbitrary constant.
This set of numbers obeys the following property,
\be\label{PropertyOriginExp}
\sum_{u\in P_K} \exp \bigl(\pm 2\pi \i (u+\delta) \bigr) \= 0 \,, \qquad K>1 \,, \quad \d \in \IR \,,
\ee
and, relatedly, 
\be\label{PropertyOrigin}
\sum_{u\in P_K} \sin \bigl(\pm 2\pi (u+\delta) \bigr) \= 0 \,, \qquad K>1 \,, \quad \d \in \IR \,.
\ee

In the main text we saw that the saddle equations for matter in the adjoint of $U(N)$ are solved by 
solutions to
\be \label{saddleAdjoint}
\sum_{k=1}^N \, \sin\bigl( 2 \pi (u_{j} - u_{k}) \bigr)  \= 0 \,, \qquad j=1,\dots,N\,.
\ee
For any~$K|N$, the configuration~$u_{j} \in P_K$, $j=1,\dots, N$, 
with an equal number~$N/K$ at each element of the set is a solution to the equations~\eqref{saddleAdjoint}.
For~$K=1$ this is because~$u_j - u_k$ vanishes for all pairs and the sine function vanishes at the origin.
For~$K>1$ we use the property~\eqref{PropertyOrigin} with~$\d=-u_j$.

For generic quivers of the type discussed in Section~\ref{Sec:SaddlePoints}, we have to analyze the 
equations given by~\eqref{gjquiver}, i.e., for~$a=1,\dots ,\nu$, $i=1,\dots , N_a$,
\be\begin{split} \label{gjquiveragain}
0  & \= \t_{aa} \, \rme^{2\pi \i \d_{aa}} \, \sum_{j=1}^{N_a} \, \sin({2 \pi  (u^a_i-u^a_j)}) \; + \\
&  \qquad \qquad + \; \sum_{b\neq a} \,  \biggl( \, \t_{ab} \, \rme^{2\pi \i \d_{ab}} \, \sum_{j=1}^{N_b} \,   \rme^{2 \pi \i (u^a_i-u^b_j) } 
\; -  \; \t_{ba} \, \rme^{2\pi \i \d_{ba}} \,  \sum_{j=1}^{N_b} \,   \rme^{2 \pi \i  (u^b_j-u^a_i)}  \, \biggr) \,.
\end{split}
\ee
Given a set of divisors~$\{ K_a ; \, K_a | N_a\}$ and~$K_a>1$, $a=1,\dots, \nu$, the configuration in which~$u^a_{j} \in P_{K_a}$ 
with an equal number~$N_a/K_a$ at each element of the set is a solution to the equations~\eqref{gjquiveragain}.
To see this we use the property~\eqref{PropertyOrigin} with~$K_a|N_a$ and~$\d=-u^a_i$  for the first line of~\eqref{gjquiveragain}, 
and the property~\eqref{PropertyOriginExp} with~$K_b|N_b$ with~$\d=-u^b_i$ for each term separately 
of the second line of~\eqref{gjquiveragain}.
The action of these solutions can be found by plugging them into the function~$g$ defined in~\eqref{gquiver}. 

%
%\bibliography{GrowthSCFTIndex}
%\bibliographystyle{JHEP}
%
%

\providecommand{\href}[2]{#2}\begingroup\raggedright\endgroup

\end{document}